\documentclass[aps,prl,floatfix,twocolumn,superscriptaddress]{revtex4}
\usepackage{bm}
\usepackage{epsf}
\usepackage{amssymb}
\usepackage{amsmath}
\usepackage{graphicx}
\usepackage{rotating}
\usepackage{epsfig}
\usepackage{psfrag}
\usepackage{amsmath}
\usepackage{hyperref}
\usepackage{subfigure}

\usepackage{color}

\begin{document}

\renewcommand{\a}{\alpha}
\renewcommand{\b}{\beta}
\newcommand{\g}{\gamma}
\newcommand{\G}{\Gamma}
\renewcommand{\d}{\delta}
\renewcommand{\S}{\Sigma}
\newcommand{\s}{\sigma}
\newcommand{\D}{\Delta}
\renewcommand{\th}{\theta}
\newcommand{\Th}{\Theta}
\renewcommand{\o}{\omega}
\renewcommand{\O}{\Omega}
\newcommand{\e}{\epsilon}

\renewcommand{\dag}{\dagger}
\newcommand{\lb}{\label}
\newcommand{\nn}{\nonumber}
\newcommand{\see}{\rightarrow}

\newcommand{\cm}[1]{\textcolor{red}{\bf #1}}

\def\br{{\bf r}}
\newcommand{\Tr}{\mbox{Tr}}
\renewcommand{\dag}{\dagger}

\newcommand{\PD}[2]{\frac{\partial{#1}}{\partial{#2}}}
\newcommand{\DD}[2]{\frac{d{#1}}{d{#2}}}
\newcommand{\BK}[1]{\left[#1\right]}
\newcommand{\bk}[1]{\left(#1\right)}
\newcommand{\bra}[1]{\left\langle{#1}\right|}
\newcommand{\ket}[1]{\left|{#1}\right\rangle}
\newcommand{\lr}[1]{\left\langle#1\right\rangle}

\newcommand{\be}{\begin{equation}}
\newcommand{\ee}{\end{equation}}
\newcommand{\ba}{\begin{eqnarray}}
\newcommand{\ea}{\end{eqnarray}}

\hypersetup{
    bookmarks=true,         
    unicode=false,          
    pdftoolbar=true,        
    pdfmenubar=true,        
    pdffitwindow=false,     
    pdfstartview={FitH},    
    pdftitle={My title},    
    pdfauthor={Author},     
    pdfsubject={Subject},   
    pdfcreator={Creator},   
    pdfproducer={Producer}, 
    pdfkeywords={keywords}, 
    pdfnewwindow=true,      
    colorlinks=true,       
    linkcolor=red,          
    citecolor=blue,        
    filecolor=magenta,      
    urlcolor=green           
}
\newcommand {\apgt} {\ {\raise-.5ex\hbox{$\buildrel>\over\sim$}}\ }
\newcommand {\aplt} {\ {\raise-.5ex\hbox{$\buildrel<\over\sim$}}\ }

\title{Quantum oscillations in a $d$-wave vortex liquid}
\author{Sumilan Banerjee}
\affiliation{Department of Physics, The Ohio State University, Columbus, Ohio, 43210}
\author{Shizhong Zhang}
\affiliation{Department of Physics, The Ohio State University, Columbus, Ohio, 43210}
\affiliation{Department of Physics and Center of Theoretical and Computational Physics, The University of Hong Kong, Hong Kong, China}
\author{Mohit Randeria}
\affiliation{Department of Physics, The Ohio State University, Columbus, Ohio, 43210}
\date{\today}

\maketitle
{\bf The observation of quantum oscillations~\cite{Leyraud2007,Sebastian2008} in 
underdoped cuprates has generated intense debate about the nature of the field-induced resistive state 
and its implications for the `normal state' of high $T_c$ superconductors. 
Quantum oscillations suggest an underlying Fermi liquid state at high magnetic fields $H$ and low temperatures, 
in contrast with the high-temperature, zero-field pseudogap state seen in spectroscopy. 
Recent heat capacity measurements~\cite{Riggs2011} show quantum oscillations together with a
large and singular field-dependent suppression of the electronic density of states (DOS),
which suggests a resistive state that is affected by the $d$-wave superconducting gap. 
We present a theoretical analysis of the electronic excitations in a vortex-liquid state,
with short range pairing correlations in space and time, that is able to reconcile these seemingly
contradictory observations. We show that phase fluctuations  
lead to large suppression of the DOS that goes like $\sqrt{H}$ at low fields,
in addition to quantum oscillations with a period determined by a Fermi surface
reconstructed by a competing order parameter.
} 

The `normal state' of the high $T_c$ superconducting cuprates remains an enigma.
In the underdoped regime, close to the Mott insulator, a large pseudogap  ($\simeq 50$ meV) 
is observed in the electronic excitations by a variety of spectroscopic and thermodynamic
probes for $T > T_c$~\cite{Timusk1999,Kanigel2006} in the absence of a magnetic field $H$. It thus came as a great surprise
when quantum oscillations were observed~\cite{Leyraud2007,Sebastian2008}
in the low-$T$ regime, once superconductivity is destroyed by $H > H_{\rm irr}$,
the irreversibility field. Such oscillations, periodic in $1/H$, are most easily understood in terms of a Fermi liquid state.
This raises the question: what is the role of the strong correlation physics that
leads to the high-$T$, zero-field pseudogap in the low-$T$, high-field quantum oscillations? 
After all a 50 T field cannot destroy a 50 meV pseudogap.

A very important clue comes from recent electronic specific heat measurements \cite{Riggs2011}
which see quantum oscillations riding on top of a strongly suppressed DOS with a $\sqrt{H}$ 
singularity at low fields. The unusual $H$-dependence points toward a $d$-wave gap
node-like structure~\cite{Volovik1993} persisting even into the resistive state.
It is not a priori clear how to reconcile the Fermi surface probed by quantum oscillations 
with the $d$-wave gap.

Further, the large suppression of the specific heat $\gamma \simeq 5$ mJ/molK$^{2}$ at $H = 50$T
(compared with the normal state value $\gamma \simeq 18$ mJ/molK$^{2}$ at  $H\!=\!0$ \cite{Loram1993}) implies that
the one has {\em not} recovered the `normal state' at the experimentally accessible fields.  
This is consistent with observations of nonlinear diamagnetism~\cite{Li2010}
and large Nernst effect \cite{Wang2006} seen in the normal state and suggestive of 
substantial phase fluctuations above $T_c$. There is also considerable
evidence that the superconducting transition in under doped cuprates is associated
with phase-disordering~\cite{Uemura1989,Emery1994,Hetel2008} rather than a gap collapse.

In this paper we make a simple model of the electronic excitations
in a vortex liquid and show how it helps reconcile, within a single framework,
the apparently contradictory observations described above. In the vortex liquid state 
the local $d$-wave pairing amplitude is non-zero, but  superconducting
correlations are short-ranged in both space and time. Our analysis generalizes
earlier studies of the mixed state of s-wave superconductors
in refs.~\cite{Maki1991,Stephen1992,Maniv2001} in two ways -- d-wave pairing and dynamical
phase fluctuations -- both of which are very important for quantum oscillations in cuprates.
We show that the effect of phase fluctuations on the electronic self-energy
leads to quantum oscillations riding on top of a strongly suppressed DOS that goes 
like $\sqrt{H}$ at small $H$. 

However, we need more than just phase fluctuations to understand
the quantum oscillations. Their observed frequency in underdoped cuprates
(unlike that in overdoped samples \cite{Vignolle2008}) is known to be
too small to be consistent with a Luttinger Fermi surface (FS). It corresponds to
an electron-like FS with area of only about 2\% of the Brillouin zone (BZ)~\cite{Leyraud2007,Sebastian2008}.
It is now accepted that these observations necessarily imply
a FS that has been reconstructed by a (possibly field-induced)
density wave order~\cite{Millis2007,Chakravarty2008,Yao2011,Harrison2011,Sebastian2011}.
Thus, to get a complete description of the underdoped cuprate experiments, we 
incorporate {\it both} phase fluctuations {\it and} a competing order parameter at the end of the paper.

\begin{figure*}
\begin{center}
\begin{tabular}{ccc}
\includegraphics[height=4.5cm]{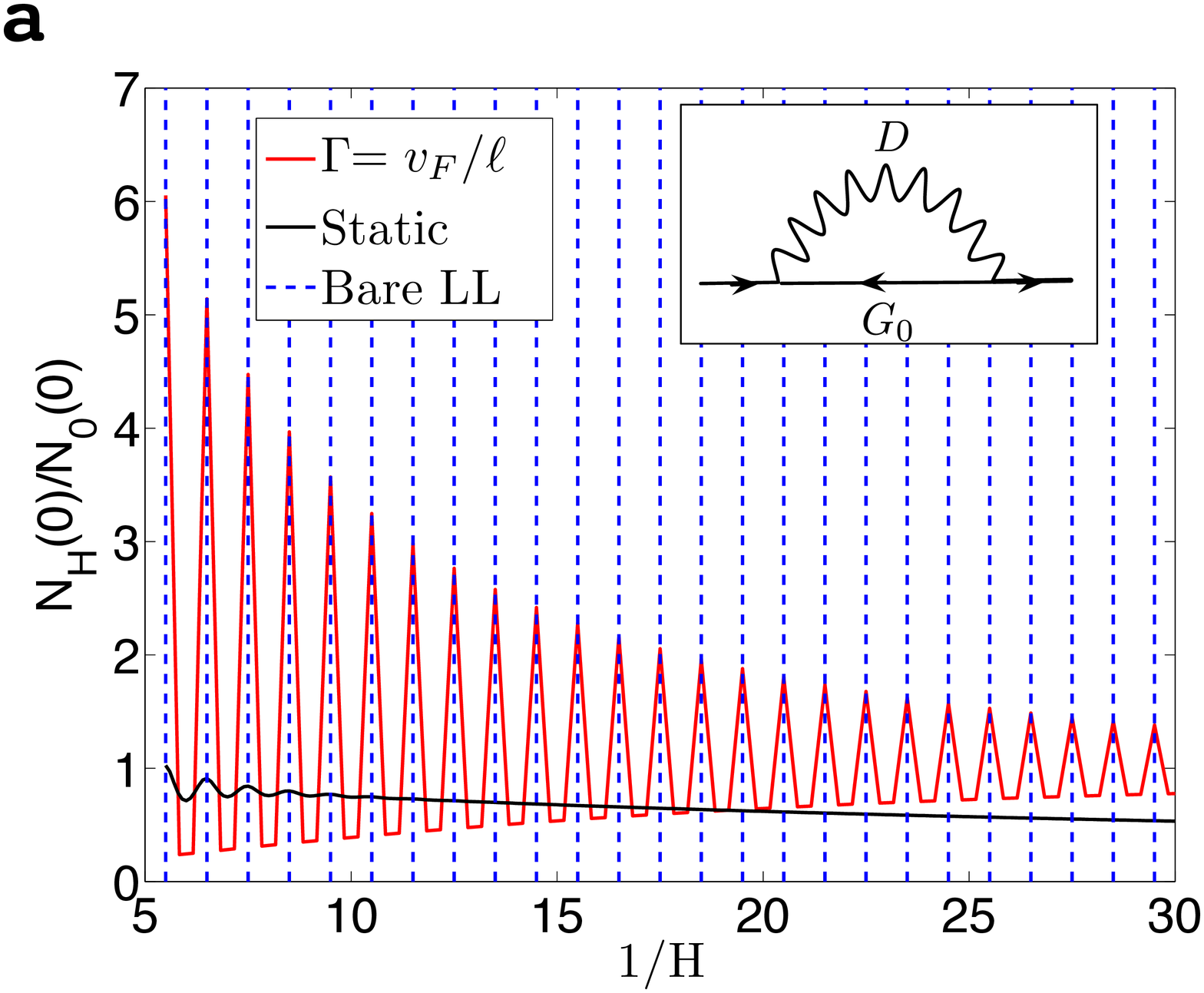}&
\includegraphics[height=4.5cm]{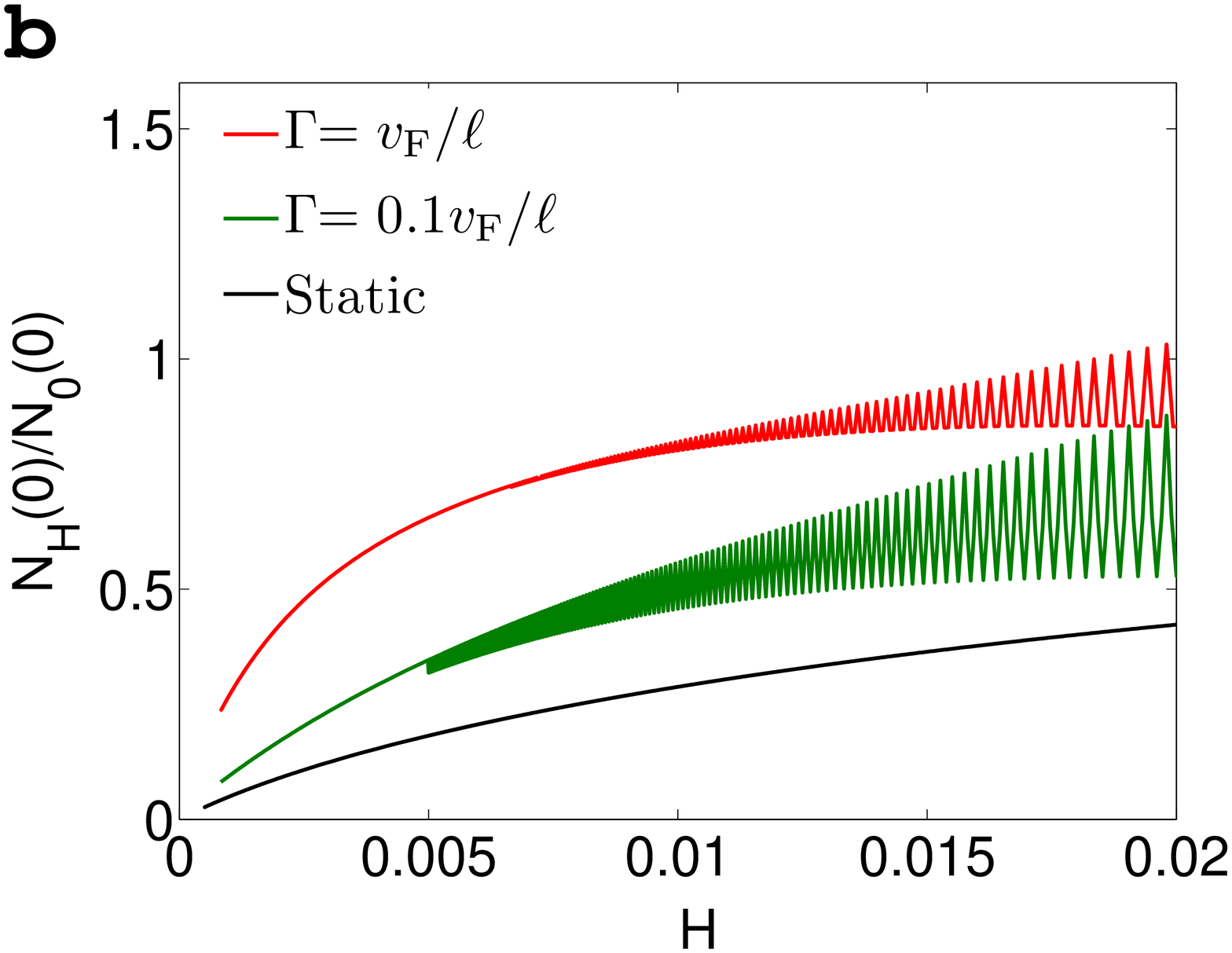}&
\includegraphics[height=4.5cm]{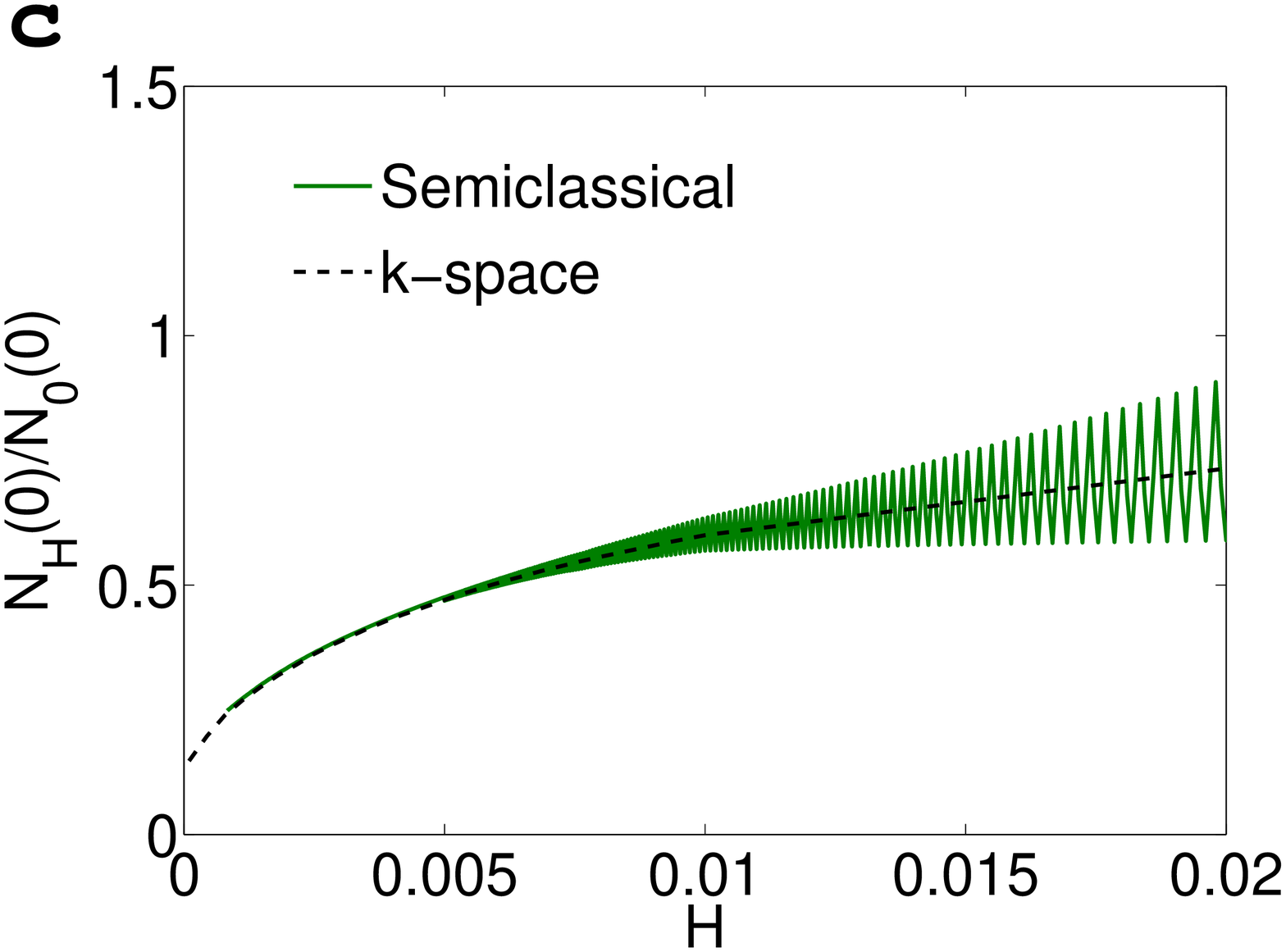}
\end{tabular}
\end{center}
\caption{{\bf Quantum Oscillations in Density of States:} DOS at the chemical potential $N_H(0)$ normalized
by the zero-field normal state $N_0(0)$. The magnetic field $H$, measured in units of $\mu$, is 
$\hbar\omega_c/\mu$ with  $\omega_c = eH/m^*c$.  
We use $\Delta\simeq 0.3$, and an impurity broadening $\gamma_0=0.01$ in units of $\mu=1$.
{\bf a}:  $N_H(0)$ vs.~$1/H$ for the static and dynamic cases ($\Gamma\neq 0$) analyzed in the Landau levels (LL) basis. 
The DOS for LL's with $\Delta=0$ is also shown. Note the $(1/H)$ periodicity of the oscillations
and the difference in the damping between the static and dynamic cases. The inset shows the self-energy approximation (see text). 
{\bf b}: $H$-dependent suppression of DOS computed in the LL basis.  We show that $N_H(0)\propto \sqrt{H}$ for small $H$.
{\bf c}:  DOS obtained from the $\mathbf{k}$-space and semiclassical quantization schemes for $\Gamma=0.1v_\mathrm{F}/\ell$.}
\label{fig.LLCalculation}
\end{figure*}

{\bf Phase Fluctuations:} 
We characterize the vortex liquid state with a simple {\it ansatz} 
for the gauge-invariant correlation function
$D_{\mu\nu}(\mathbf{r},t)\equiv\left\langle\Psi_{\mu}(\mathbf{r},t)\Psi_\nu^*(\mathbf{0},0)
\exp{(i\frac{2e}{\hbar c}\int_\mathbf{0}^\mathbf{r}\mathbf{A}\cdot d\mathbf{l})}\right\rangle$.
The field $\Psi_\mu(\mathbf{r},t)$ describes singlet pairs on the bond $(\mathbf{r},\mathbf{r}+a\hat{\mu})$
with $a$ being the Cu-Cu lattice spacing of the $\mathrm{CuO}_2$ square lattice with sites $\mathbf{r}$; $\mu=\pm \hat{x},\pm \hat{y}$
and $\mathbf{A}$ is the vector potential for the magnetic field $\mathbf{H}=H\hat{z}$.
The retarded correlation function 
\begin{equation}
D_{\mu\nu}^{(R)}(\mathbf{r},t) = s_{\mu\nu}\Delta^2 \exp{(-r^2/2\ell^2)}\exp{(-\Gamma t)}\Theta(t) \label{eq.PairPropagator}
\end{equation}
is assumed to be {\it short ranged in space and time}; $\Theta(t)$ is the Heaviside step function.  
$\Delta$ is the {\em local} d-wave pairing amplitude that persists
above $H_{\rm irr}$, as expected for a superconductor where the resistive transition is governed by phase fluctuations.
The d-wave nature is described by $s_{\mu\nu}=1$ for $\mu=\pm \nu$ and $-1$ otherwise.
The spatial decay in (\ref{eq.PairPropagator}) is on the magnetic length scale $\ell=\sqrt{\hbar c/eH}$, 
set by the average inter-vortex separation in the extreme type-II limit. We work in a regime where the cyclotron radius 
$R_c \gg \ell \gg \xi_0 \gtrsim k_F^{-1}$, where $\xi_0$ is the vortex core radius and $k_F^{-1}$ the interparticle spacing.

The temporal decay in (\ref{eq.PairPropagator}) is governed by an energy scale $\hbar\Gamma$ that characterizes vortex motion. 
On general grounds we expect $0 < \Gamma \leq v_\mathrm{F}/\ell$. 
The upper limit arises from  ballistic motion of vortices with the Fermi velocity $v_\mathrm{F}$.
For simplicity, we write $\Gamma = \alpha v_{\rm F}/\ell$ with $0< \alpha \leq1$ in the {\it dynamic} case where vortices are mobile.
The {\it separable form} (\ref{eq.PairPropagator}) simplifies the algebra,
but a more elaborate non-separable correlator with $\Gamma = Dq^2$, where $D$ is the vortex diffusion coefficient,
is not expected to change our conclusions qualitatively~ \cite{Senthil2009,Micklitz2009}.
We will also find it useful to compare our results for dynamic phase fluctuations ($\Gamma \neq 0$) with the
the {\it static} case $D_{\mu\nu}(\mathbf{r},t) = s_{\mu\nu}\Delta^2 \exp{(-r^2/2\ell^2)}$ with
time-independent phase fluctuations~\cite{Maki1991,Stephen1992,Maniv2001}. 

We use the simplest approximation for self-energy $\Sigma({\bf k},\omega)$ (inset of Fig.\ref{fig.LLCalculation} {\bf a})
to find the effect of phase fluctuations on electronic excitations in the vortex-liquid state. Our approach generalizes
the static, s-wave analysis of refs.~\cite{Maki1991,Stephen1992}.
A similar self-energy has also been used for the pseudogap phase 
of cuprates \cite{Senthil2009,Micklitz2009,Banerjee2011}, but no calculations have been presented for quantum oscillations.

The central quantity of interest to understand quantum oscillations
is the single-particle DOS at the chemical potential $N(0)$ at $T=0$
as a function of the external field $H$.
We use the self energy $\Sigma$ to compute the electronic Green's function 
$G = 1/[G_0^{-1}-\Sigma]$, where $G_0$ is the free Green's function.
The imaginary part of $G$ then gives us the DOS $N(\omega)$.
We note that there is no anomalous part of the Green's function,
since $\langle \Psi_\mu(\mathbf{r},t)\rangle=0$ in the  absence of long-range phase coherence.
We will focus first on the simple case of parabolic dispersion, where we can do the calculation in two different 
ways: in the Landau level (LL) basis and in momentum (${\bf k}$) space. We then use the ${\bf k}$-space approach to
shed light on the crucial role of the dynamics of phase fluctuations.  Finally we examine quantum oscillations for arbitrary dispersion
using semi-classical quantization of ${\bf k}$-space results.

\begin{figure}
\begin{center}
\begin{tabular}{ccc}
\includegraphics[height=3.3cm]{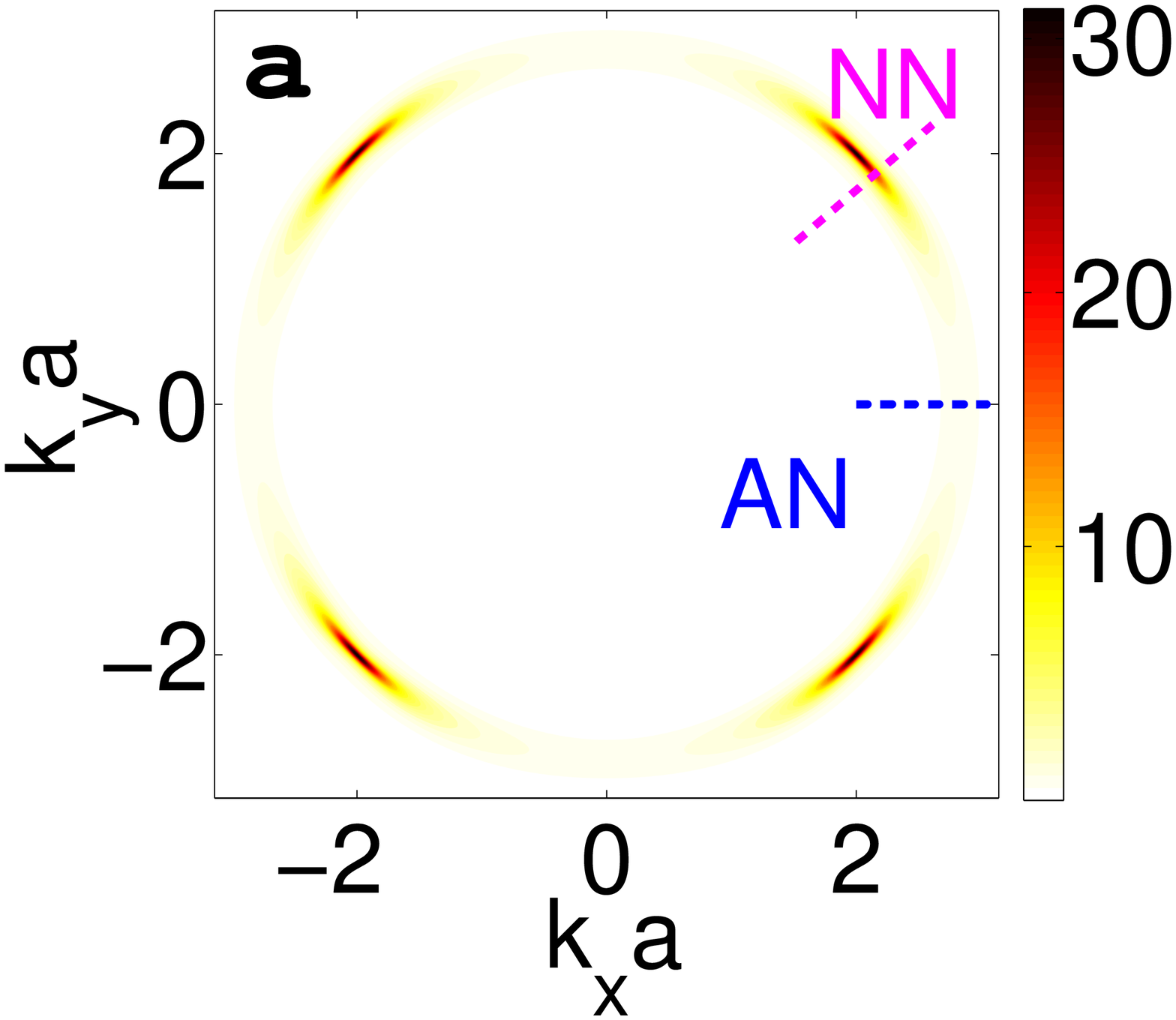}&
\includegraphics[height=3.3cm]{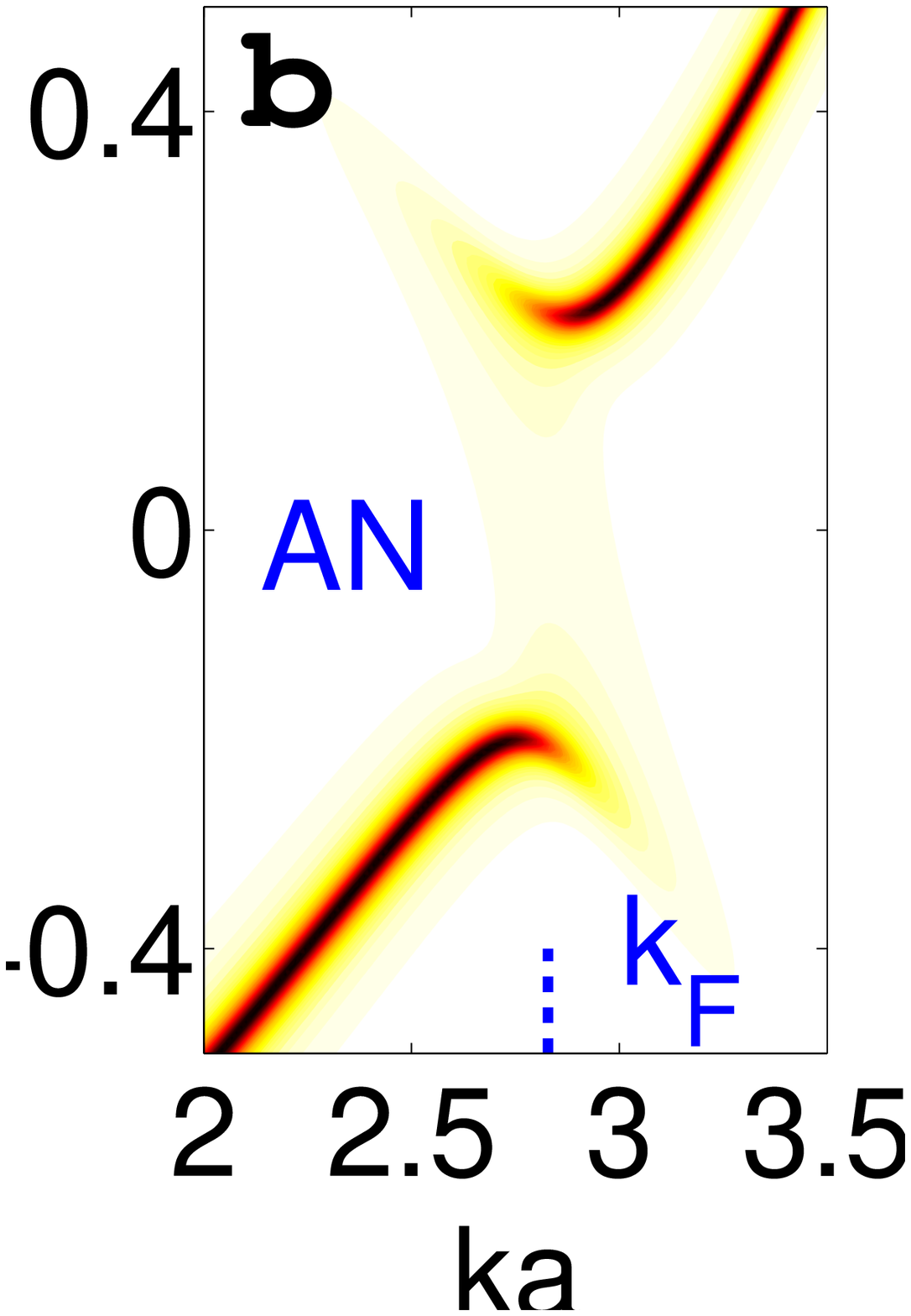}&
\includegraphics[height=3.3cm]{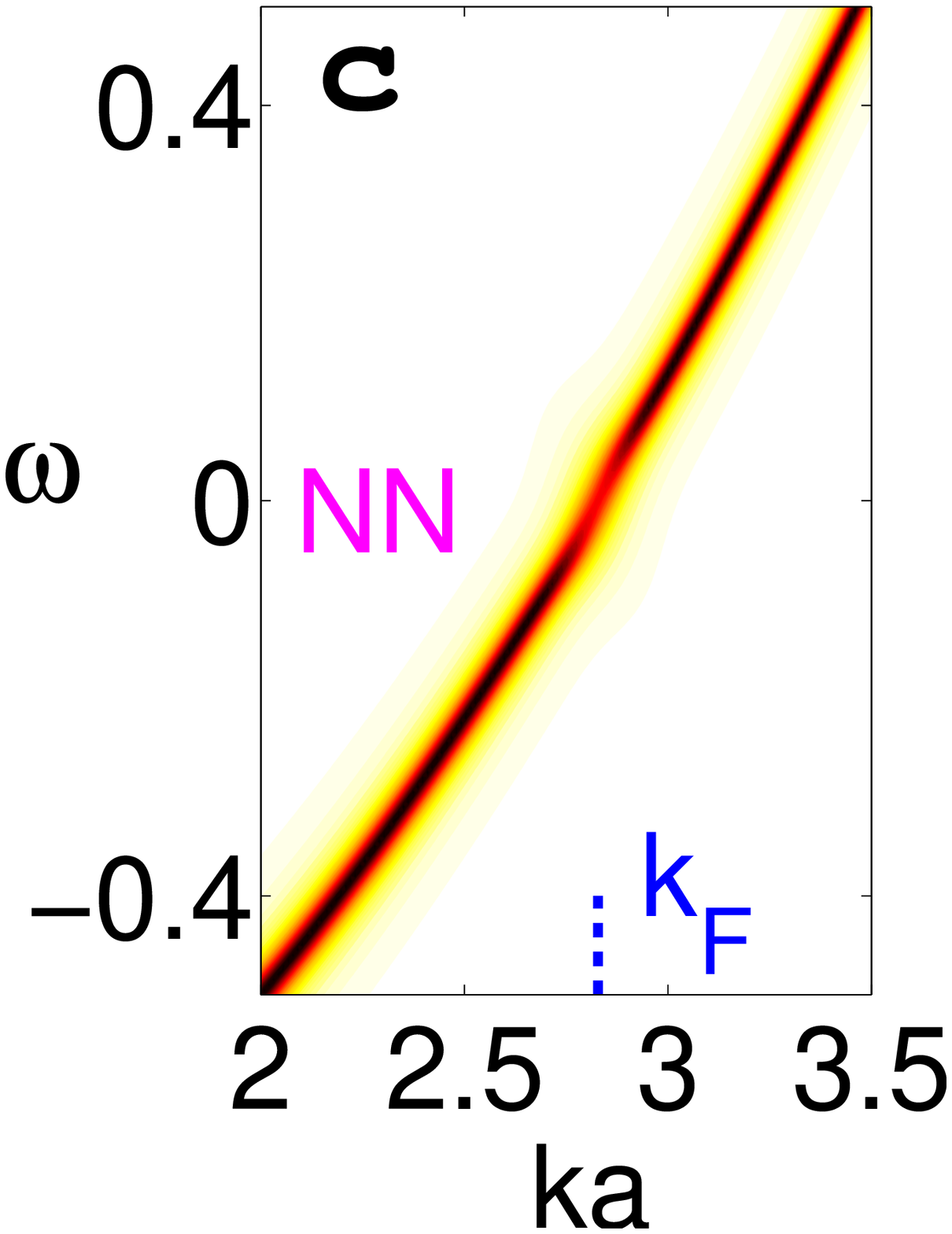}\\
\includegraphics[height=3.3cm]{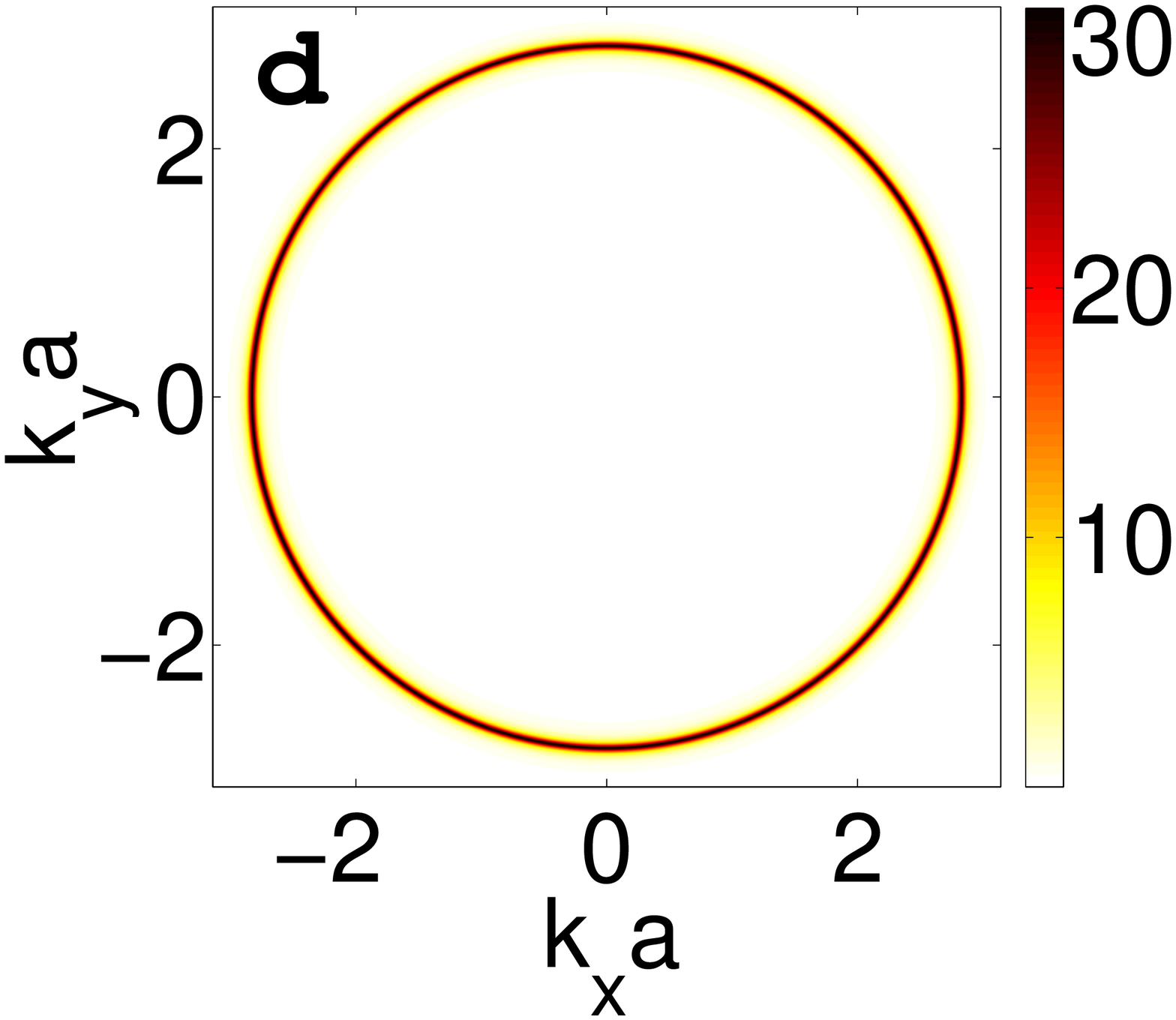}&
\includegraphics[height=3.3cm]{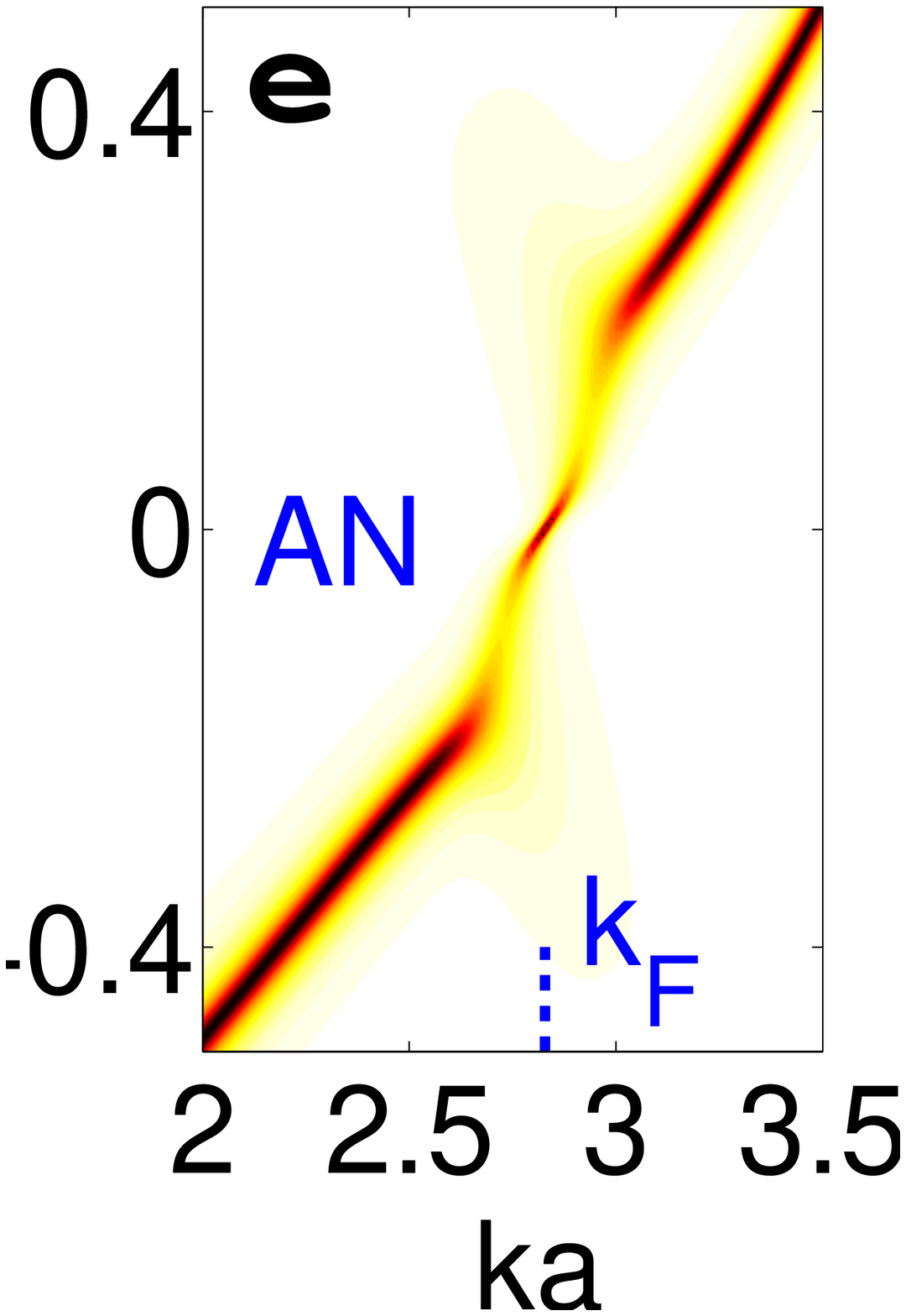}&
\includegraphics[height=3.3cm]{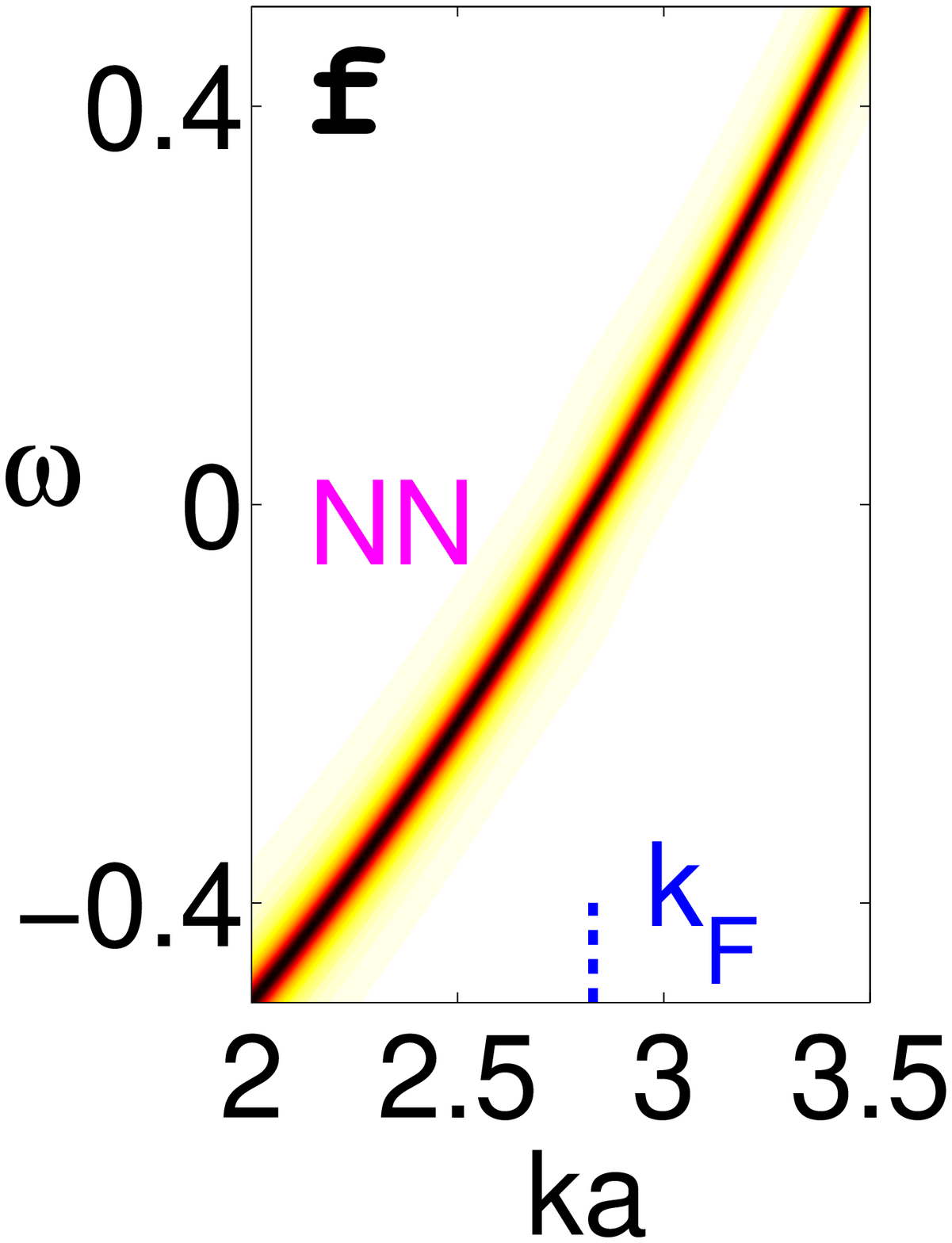}
\end{tabular}
\end{center}
\caption{{\bf Spectral functions}: False color plots of $A(\mathbf{k},\omega)$.
{\bf a} $A(\mathbf{k},\omega=0)$, for static phase fluctuations
contributing to the self-energy $\Sigma(\mathbf{k},0)$, showing intensity along Fermi arcs near the nodes. 
We use a parabolic dispersion with $H=0.003$ (as in Fig.~\ref{fig.LLCalculation}, we express $H$ in units of $\mu=1$ and $\Delta\simeq 0.3$). 
{\bf b}: $\omega$ vs. ${\bf k}$ dispersion near the antinode (AN) along a cut perpendicular to the FS, showing a Bogoliubov-like dispersion with a AN pseudogap.
{\bf c}: Gapless dispersion near the node (NN). We show in panel ({\bf a}) both the AN and NN momentum cuts.
{\bf d}: $A(\mathbf{k},\omega=0)$, for dynamic phase fluctuations (with $\Gamma=v_\mathrm{F}/\ell$), shows the
restoration of the full Fermi surface.
{\bf e}, {\bf f}: Similar to plots ({\bf c,d}) for the dynamical case. Note particularly the appearance of a zero-energy
quasiparticle inside the pseudogap in the AN cut (panel {\bf e}) due to quantum motional narrowing (see text).
} 
\label{fig.kSpaceCalculation}
\end{figure}
 
{\bf Landau level analysis:} Consider electrons with a dispersion $\epsilon_\mathbf{k}=\hbar^2\mathbf{k}^2/2m^*$
and chemical potential $\mu$ in a magnetic field $H$.
In the LL basis $G_0(n,i\omega_l)=(i\omega_l-\xi_n)^{-1}$, where
$\omega_l=(2l+1)\pi T$ is the fermionic Matsubara frequency and the spectrum
$\xi_n=(n+1/2)\hbar \omega_c-\mu$, with cyclotron frequency $\omega_c=(eH/m^*c)$
and LL index $n$. Using this $G_0$ and the fluctuation propagator  $D_{\mu\nu}$
we obtain the self-energy (see Methods) making the approximation of retaining only diagonal terms in the LL self-energy matrix $\Sigma(n,n';i\omega_l)$.
We then calculate the DOS 
$N(\omega)=-(1/\pi\ell^2)\sum_n \mathrm{Im}G(n,i\omega_l\rightarrow\omega+i0^{+})$ with $\omega$  measured from $\mu$. 
All results shown here are for $T=0$. 

We show in Fig.\ref{fig.LLCalculation}({\bf a}) and ({\bf b}) the DOS $N_H(0)$ in a field $H$, 
normalized to the $H=0$ normal state $N_0(0)$. Both curves in panel ({\bf a}) 
exhibit quantum oscillations periodic in $(1/H)$ with the usual
Onsager frequency given by the Fermi surface area. 
The index along $x$-axis directly counts the number of filled LL $n_\mathrm{F}\sim \mu/\hbar \omega_c$. 
There is however a striking difference between the damping of the
static (black curve) and dynamic (red curve) results
in Fig.\ref{fig.LLCalculation}({\bf a}). In the static case, the oscillations decay rapidly as
$\exp{(- \pi/\tau\omega_c)}$ from the large {\it intrinsic} damping 
$1/\tau\sim |\mathrm{Im}\Sigma(n_\mathrm{F},0)|\neq 0$ arising from scattering of electrons 
from static phase fluctuations. (We note that the static $d$-wave oscillations, though strongly damped,
are still are much less so than the static $s$-wave results~\cite{Maki1991,Stephen1992} due to
the averaging over the sign changes in the local order parameter.)
In contrast, there is no intrinsic damping in the dynamic case, with
$\mathrm{Im}\Sigma(n_\mathrm{F},0) =0$.
The damping in Fig.\ref{fig.LLCalculation}({\bf a}) arises from a 
small impurity broadening $1/\tau\sim\gamma_0$ put in by hand in $G_0$, and
inevitably present in real materials. Below, we will gain insight into why the intrinsic damping
due to dynamic phase fluctuations vanishes.   
This has direct implication for quantum oscillations 
in cuprates, where no perceptible damping, in addition to that expected from impurity scattering, 
has been observed.
    
In Fig.~\ref{fig.LLCalculation}({\bf b}), we plot $N_H(0)/N_0(0)$ as a function of $H$ (rather than $1/H$)
for low fields with $n_\mathrm{F}$ of order hundred.  
Quantum oscillations are seen only for $\Gamma \neq 0$ and completely suppressed
for the static case. 
We also see a large, $H$-dependent DOS suppression relative to the zero-field normal state, 
with, as we show below, a $\sqrt{H}$ singularity at small $H$.
Quantitatively, the suppression depends  on $\Gamma$, becoming larger with
decreasing $\Gamma$ and most pronounced in the static case. The DOS suppression also depends on 
the pairing strength $\Delta$, which we take to be $H$-independent, as is reasonable for low fields.    
 
{\bf Momentum space analysis:} 
To gain insight into the LL results, we turn to a ${\bf k}$-space analysis.  The self-energy $\Sigma(\mathbf{k},i\omega_l)$ 
is calculated using the fluctuation propagator (\ref{eq.PairPropagator}) with 
$G_0(\mathbf{k},i\omega_l)=(i\omega_l-\epsilon_\mathbf{k}+\mu)^{-1}$;  (see Methods for details).
We first look at $\epsilon_\mathbf{k}=\hbar^2\mathbf{k}^2/2m^*$ and then generalize to arbitrary dispersion later.

It is helpful to look at the one-electron spectral function $A(\mathbf{k},\omega)=-(1/\pi)\mathrm{Im}G^{(R)}(\mathbf{k},\omega)$.
We show that dynamical phase fluctuations restore a zero-energy quasiparticle~\cite{Senthil2009,Micklitz2009}
at the antinodal $k_F$. We may think of this as quantum {\it motional narrowing}, with the effect of pairing on the
spectral function washed out on the longest time scales, as we now describe in detail. 

\begin{figure}
\begin{center}
\begin{tabular}{cc}
\includegraphics[height=4cm]{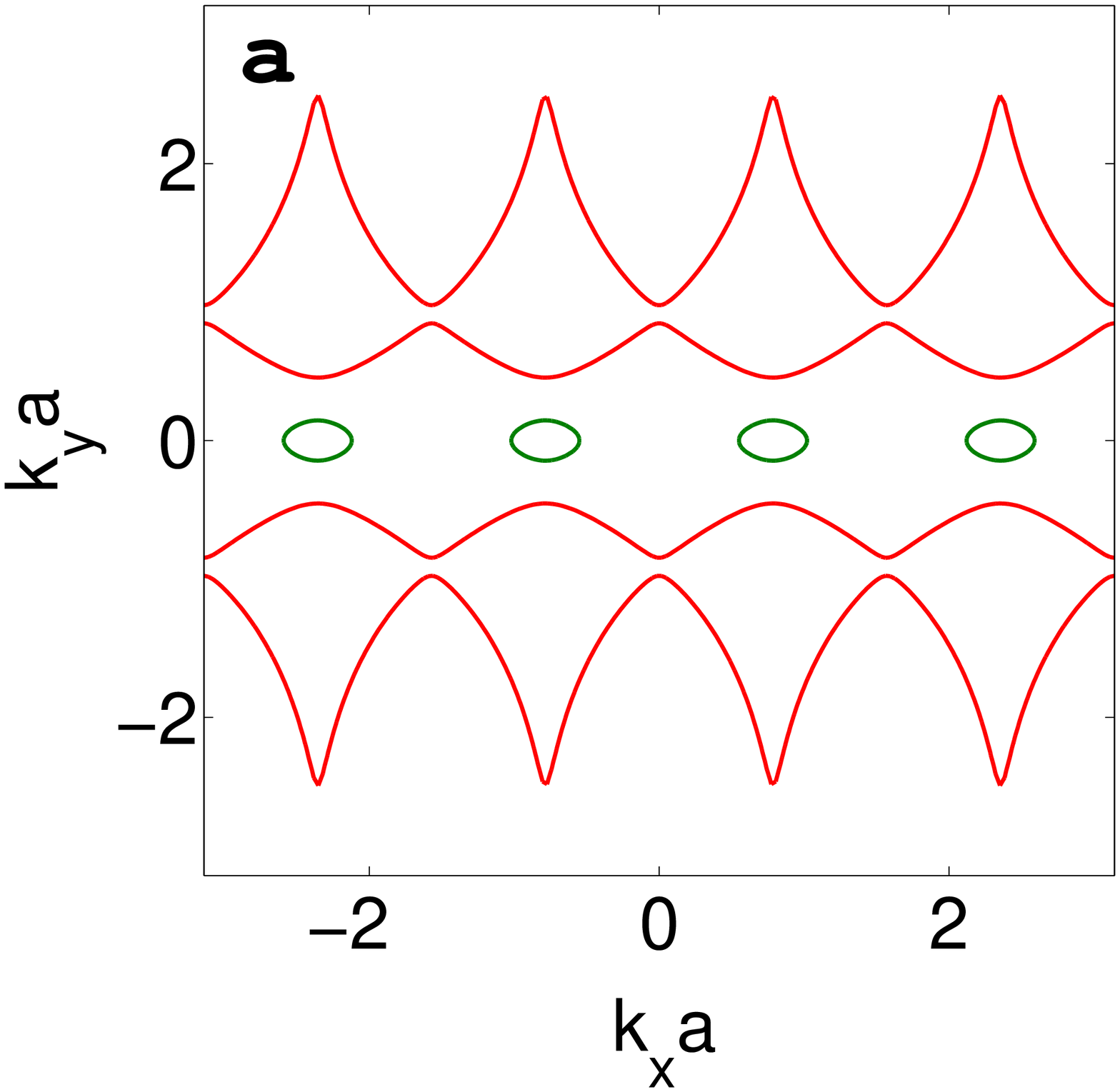}&
\includegraphics[height=4cm]{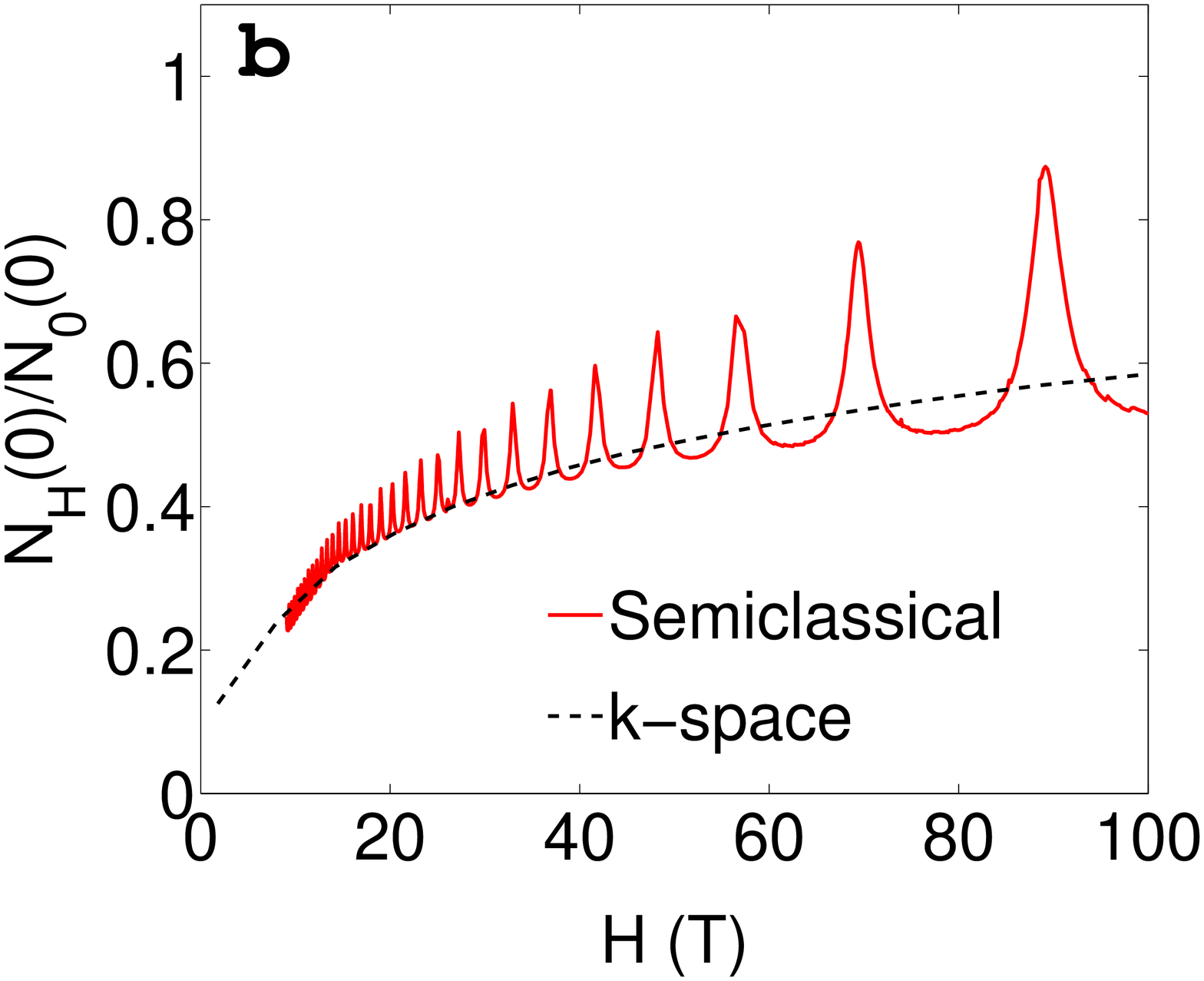}
\end{tabular}
\end{center}
\caption{{\bf Fermi surface reconstruction} {\bf a}: Open and closed FS segments shown in the original 
Brillouin zone (BZ ) for density wave order scenario of ref.~\cite{Yao2011}. We choose parameters
$t'/t=-0.4$, $\phi_N=0.143$, $V=0.11 t$ and $V'=0.09t$ (enhanced  $V$'s to make the reconstruction visible)
with $\mu=-1.05t$. 
{\bf b}: DOS quantum oscillations obtained from semiclassical quantization of the reconstructed FS in a vortex liquid state.
Here $t'/t=-0.4$, $\phi_N=0.143$, $V=0.02 t$, $V'=0.01t$ with $\mu=-1.05t$, so that the hole doping 
$x\approx 0.12$ and the closed orbit area $\simeq$ 2\% of the BZ. The DOS also includes the contributions of open orbits.
The $H$-dependent suppression of the DOS arises from phase fluctuations with
$\Delta\simeq 0.3t$ and $\Gamma=0.1v_\mathrm{F}/\ell$. Here $\gamma_0=0.01t$.   
}
\label{fig.FSReconstruction}
\end{figure}

We plot in Fig.~\ref{fig.kSpaceCalculation} ({\bf a,b,c}) the spectral functions for 
static case with time-independent phase fluctuations,
that should be contrasted with the corresponding results in panels ({\bf d,e,f}) for the dynamic case
with $\Gamma\neq 0$.  In the static case, we see in panels ({\bf a,b,c}) that 
phase fluctuations broaden the node of the d-wave SC into ``Fermi arc", a region of gapless excitations where
the spatial fluctuation-induced line width $v_F/\ell$ overwhelms the gap 
$\Delta_{\bf k} = \Delta[\cos(k_x a) - \cos(k_y a)]/2$.  There is a pseudogap in the antinodal region
where $|\Delta_{\bf k}| > v_F/\ell$ as seen in both panels . 
We note that the static case is {\it not} the (highly singular) limit $\Gamma \rightarrow 0$ at $T=0$, 
but more closely related to the high temperature regime where $T > \Gamma$; see Methods.
In the high-$T$ regime, the phase fluctuations are classical~\cite{Banerjee2011} and we can ignore their
time dependence. Thus the physics of the static results is relevant for high-$T$ experiments 
like angle-resolved photoemission spectroscopy (ARPES). On the other hand, the quantum 
oscillation experiments are in the very different low-$T$ limit, where the dynamics of phase fluctuations 
cannot be ignored.    

The results with $\Gamma \neq 0$ dynamics are qualitatively different from the static case. 
We see from Fig.~\ref{fig.kSpaceCalculation} ({\bf d}) that one recovers the full Fermi surface (FS),
albeit with a highly anisotropic self-energy, as illustrated by the $A(\mathbf{k},\omega)$
dispersion plots in panels ({\bf e,f}), along two representative 
momentum cuts perpendicular to the FS, one near the antinode (AN) in
the SC state and the other near the node (NN). We can understand the 
appearance of the zero-energy quasiparticle at the AN by looking at the self-energy.
In contrast to the static case, which has a large antinodal $|\mathrm{Im}\Sigma({\bf k}_{\rm F},\omega)|$ at low energies, dynamical phase fluctuations lead 
to $|\mathrm{Im}\Sigma(\mathbf{k}_{\rm F},\omega)| \sim \omega^2$ for $|\omega| \ll \Gamma$,
the quantum motional narrowing mentioned above.
The corresponding $\mathrm{Im}\Sigma(\mathbf{k}_{\rm F},\omega)$ then leads to a 
quasiparticle pole at the chemical potential, even though the self-energy effects are
are strongly $\mathbf{k}$-dependent as seen from Fig.~\ref{fig.kSpaceCalculation}({\bf e,f}).

The existence of sharp quasiparticles all around the full FS immediately leads to
the quantum oscillations with the Onsager frequency.
We define a {\it renormalized} dispersion $\widetilde{\epsilon}_\mathbf{k}=\epsilon_\mathbf{k}+\Sigma(\mathbf{k},0)$ 
for low-energy quasiparticles, which has a non trivial $H$-dependence from the self-energy. 
We then use a semiclassical prescription~\cite{Luttinger1961} to quantize the
orbits (see Methods). The resulting DOS from this $\mathbf{k}$-space analysis is shown in 
Fig.~\ref{fig.LLCalculation}({\bf c}), with a small impurity scattering $\gamma_0$ that damps the quantum oscillations.

The most non-trivial aspect of this result is that the quantum oscillations ride on top of a 
large, field-dependent suppression of the DOS $N_H(0)$, just as we saw in the LL analysis (Fig.\ref{fig.LLCalculation}({\bf b})). 
We can analyze this suppression by looking at the ``average'' DOS (without any semiclassical quantization), shown as the dashed curve in Fig.~\ref{fig.LLCalculation}({\bf c}).
We can show analytically that the $H$-dependent self-energy
$\Sigma(\mathbf{k},0)$ leads to $N_H(0)\propto \sqrt{H}$ as $H\rightarrow 0$ (see Methods)
in the static case. This reproduces the celebrated Volovik result \cite{Volovik1993} from quite a different route.
The residual value of $N_H(0)$ at $H=0$ is due to the impurity scattering $\gamma_0$.     
 
{\bf Fermi surface reconstruction by a competing order:} 
The analysis presented above shows that while phase fluctuations are able to reconcile quantum oscillations
with a large suppression of the DOS that goes like $\sqrt{H}$, they do not affect the oscillation frequency.
Thus, to get a complete description of the underdoped cuprate experiments, we need to 
incorporate {\it both} phase fluctuations {\it and} a competing order.
We can incorporate any one of the proposed broken symmetries~\cite{Millis2007,Chakravarty2008,Yao2011,Harrison2011,Sebastian2011}
within our  ${\bf k}$-space formulation. Only experiments will decide which competing order is most relevant for
a particular material. 
Once SC long-range order is destroyed by the field, it is natural that the ground state of a 
lightly doped Mott insulator exhibits a new density wave instability that reconstructs the FS.
However, we are firmly of the opinion that the large ($\simeq 50$meV) AN pseudogap cannot arise from a 
small (or subtle) symmetry breaking, and it is not reasonable to use a large symmetry breaking potential
to reconstruct the FS.

To understand the interplay of FS reconstruction and phase fluctuations,
we analyze, as an illustrative example, the density-wave order proposed in ref.~\cite{Yao2011} 
with $\mathbf{Q}=(\frac{\pi}{2},0)$ consistent with recent high-field NMR data~\cite{Wu2011} 
Following ref.~\cite{Yao2011}, we start with 
$\e_{\bf k}=-2t[(1+\phi_N)\cos{k_xa}+(1-\phi_N)\cos{k_ya}]-4t'\cos{k_x a}\cos{k_ya}$,
where $t$ ($t'$) is the nearest (next-nearest) neighbour hopping and $\phi_N$  
the nematicity.  Phase fluctuations renormalize this dispersion to 
$\widetilde{\e}_{\bf k}={\e}_{\bf k}+\S({\bf k},0)$, with the self-energy discussed above. 
To find the FS reconstruction, we diagonalize the Hamiltonian 
\begin{displaymath}
\mathcal{H}_{\bf k} =
\left( \begin{array}{cccc}
\widetilde{\e}_{\bf k} & V & V' & V \\
V& \widetilde{\e}_{{\bf k}+{\bf Q}} & V & V' \\
V' & V & \widetilde{\e}_{{\bf k}+2{\bf Q}} & V\\
V & V' & V &\widetilde{\e}_{{\bf k}+3{\bf Q}}
\end{array} \right),
\end{displaymath}
where $V$ and $V'$ are the density wave potentials.

We semiclassically quantize the resulting energy dispersion. The results are shown in Fig.\ref{fig.FSReconstruction}. The chemical potential is fixed for $x\approx 0.12$. As shown in Fig.\ref{fig.FSReconstruction} {\bf a}, there are small electron-like FS pockets and open FS sheets. The pockets have an area around $2\%$ of the original BZ and there is only one pocket in the new BZ. Although the open FS sheets do not contribute to oscillations of DOS, they give rise to large contribution to DOS in the absence of pairing. In a vortex liquid these DOS from the open orbits are largely suppressed as shown in Fig.\ref{fig.FSReconstruction} {\bf b}. Hence in this case the frequency as well as the suppression of DOS are both in good agreement with experiments \cite{Leyraud2007,Sebastian2008,Riggs2011}.  

In conclusion, we have presented a calculation of the electronic excitations in a field-induced vortex liquid state with 
short-ranged d-wave pairing fluctuations. This leads to a non-trivial self energy that is responsible for
a large suppression of the DOS, with a constant plus $\sqrt{H}$ variation at low fields.
The dynamics of phase fluctuations restores the zero-energy quasiparticles via quantum motional narrowing
at low temperatures, thus leading to the full FS, absent in the zero-field, high-$T$ normal state. 
A competing order parameter that breaks translational symmetry then
reconstructs the FS to give the observed low frequency of the quantum oscillations.   

{\it Acknowledgments}: We gratefully acknowledge stimulating conversations with Steve Kivelson, Mike Norman 
and Suchitra Sebastian, and the support of DOE-BES DE-SC0005035.

\section{Methods}

{\em Dynamics of Fluctuations:} 
We characterized the vortex liquid state by the ansatz \eqref{eq.PairPropagator} that describes short ranged d-wave phase
fluctuations. The gaussian spatial decay is motivated by Ginzburg-Landau theory. The dynamics of the fluctuations 
has the form $1/(|\omega_l| + \Gamma)$ in Matsubara space. The static
approximation, discussed in the text and used previously for s-wave SC's~\cite{Maki1991,Stephen1992}, corresponds to 
$(1/T)\delta_{\omega_l,0}$. The static case is thus related to the high temperature regime $T \gg \Gamma$,
and {\em not} the (highly singular) $\Gamma \rightarrow 0$ limit of the dynamics case. 
At high $T$, the classical, thermal fluctuations that dominate are time-independent.

$\mathbf{k}$-{\em space analysis:} 
We find the self-energy in the static case
\begin{equation}
\begin{split}
\Sigma(\mathbf{k},\omega)&=\sqrt{2}\frac{\Delta_\mathbf{k}^2\ell}{v_\mathbf{k}}F_\mathrm{D}\left(\frac{(\omega+\xi_\mathbf{k})\ell}{\sqrt{2}v_\mathbf{k}}\right)\\
&-i\sqrt{\frac{\pi}{2}}\frac{\Delta_\mathbf{k}^2\ell}{v_\mathbf{k}}\exp{\left(-\frac{(\omega+\xi_\mathbf{k})^2\ell^2}{2v_\mathbf{k}^2}\right)},
\label{k-space-self-energy}
\end{split}
\end{equation}
where $F_\mathrm{D}(x)=-i\sqrt{\pi} \mathrm{erf}(ix)/2$ is the Dawson function. 
Here $\Delta_\mathbf{k}=\Delta(\cos{k_xa}-\cos{k_ya})/2$,
$\xi_\mathbf{k} = \epsilon_\mathbf{k} - \mu$ and
$v_\mathbf{k}=\left|\partial \xi_\mathbf{k}/\partial \mathbf{k}\right|$. Derivations are sketched in the Supplementary Information.

In the dynamic case, the imaginary part of $\Sigma$ is
\begin{equation}
\begin{split}
&\mathrm{Im}\Sigma(\mathbf{k},\omega)=-\frac{\Delta_\mathbf{k}^2\ell}{2\sqrt{2\pi}v_\mathbf{k}}
\int d\Omega\Big[\coth{\Big(\frac{\Omega}{2T}\Big)}\label{dynamic.im.sigma}\\
&-\tanh{\Big(\frac{\Omega-\omega}{2T}\Big)}\Big]\frac{\Omega}{\Omega^2+\Gamma^2}
\exp\Big[{-\frac{(\omega-\Omega+\xi_\mathbf{k})^2\ell^2}{2v_\mathrm{k}^2}}\Big]
\end{split}
\end{equation} 
We take the $T=0$ limit of this result and obtain the
real part by the Kramers-Kronig transform of (\ref{dynamic.im.sigma}).

$\sqrt{H}$-{\em behavior:}  
At sufficiently low fields ($\ell\rightarrow \infty$) we may use the static approximation.
As $H\to 0$, the static self-energy (\ref{k-space-self-energy}) reduces to the $d$-wave BCS-Gor'kov form
$\Sigma(\mathbf{k},i\omega_l)=\Delta_\mathbf{k}^2/(i\omega_l+\xi_\mathbf{k})$. For small $H$ we expand 
$\Sigma(\mathbf{k},\omega=0)$ in $\mathbf{p}=\mathbf{k}-\mathbf{k}_N$ around one of the four nodes at $\mathbf{k}_N$,
with $p_\parallel=\mathbf{p}.\hat{v}_\Delta$ and $p_\perp=\mathbf{p}.\hat{v}_\mathrm{F}$. 
We get $\mathrm{Re}\Sigma(\mathbf{k},0)\approx (v_\Delta^2/v_\mathrm{F})p_\parallel^2p_\perp\ell^2$ and 
$\mathrm{Im}\Sigma(\mathbf{k},0)\approx -\sqrt{\frac{\pi}{2}}(v_\Delta^2/v_\mathrm{F})p_\parallel^2\ell$,
where $v_\Delta=\left|\frac{\partial \Delta_\mathbf{k}}{\partial \mathbf{k}}\right|$ and $v_\mathrm{F}$ are evaluated at $\mathbf{k}_N$.
Using this field dependence of $\Sigma(\mathbf{k},0)$ in the spectral function
leads to $N_H(\omega=0)\propto \sqrt{H}$ as $H \rightarrow 0$, thus
recovering Volovik's result \cite{Volovik1993}.

We have also generalized the above asymptotic analysis to include impurity scattering $\gamma_0$.
To make the algebra tractable, we use a Lorentzian form for $\mathrm{Im}\Sigma$, rather than the Gaussian in 
\eqref{k-space-self-energy}. We then find a result of the form 
$N_H(0)= A\gamma_0 + B\sqrt{H}$ as $H \rightarrow 0$,
in excellent agreement with the numerical results shown in the text. 

{\em Semiclassical quantization:} 
We use the semiclassical prescription for quantizing electron orbits with the
renormalized dispersion $\widetilde{\epsilon}_\mathbf{k} = \epsilon_\mathbf{k}+\Sigma(\mathbf{k},0)$.
For a given field $H$, we generate a set of energy levels
$\{\widetilde{\epsilon}_n\}$ as the solutions of 
$\mathcal{A}(\widetilde{\epsilon}_n)\simeq(2\pi n/\ell^2)$, where 
$\mathcal{A}(\widetilde{\epsilon})$ is the $\mathbf{k}$-space area enclosed by a closed orbit at energy $\widetilde{\epsilon}$
and $n$ is a positive integer. We then use the $\widetilde{\epsilon}_n$'s to compute the DOS.

{\em LL Analysis:}   
The $d$-wave self-energy $\Sigma(n,n';i\omega_l)$ is a 
matrix in the LL basis, in contrast with the $s$-wave case~\cite{Stephen1992}.
For large LL index $n$, the matrix elements decay like $\exp[-(n-n')^2/4n]$ away from the diagonal.
In the spirit of our calculation, which does not treat fluctuations self-consistently,
we neglect off-diagonal terms and retain only the diagonal ones, denoted by $\Sigma(n,i\omega_l)$.
For the static case, the self-energy at large $n$ is given by
\begin{equation}
\begin{split}
&\Sigma(n,\omega)=\frac{\Delta^2}{4}\sum_{n_1}\frac{I_{nn_1}}{\omega+\xi_{n_1}+i0^{+}}\\
&\times\left[1+J_0\left(2\sqrt{n+n_1}\frac{a}{\ell}\right)-2J_0\left(\sqrt{2(n+n_1)}\frac{a}{\ell}\right)\right],
\end{split}
\end{equation} 
where $I_{nn_1}=(n+n_1)!/n!n_1!2^{n+n_1+1}$ and $J_0(x)$ the zeroth Bessel function.

\begin{widetext}
\section{Supplementary information}
\subsection{Electronic Green's function in a $d$-wave vortex liquid}
Here we give some technical details of the calculation of electronic self-energy and Green's function
in a vortex liquid.
The $d$-wave pairing is described by the field $\Psi_\mu(\mathbf{r}+ a\hat{\mu}/2,\tau)$ defined on 
the link $(\mathbf{r},\mathbf{r}+a\hat{\mu})$. Here $\tau$ is the imaginary time with $0\leq \tau \leq \beta\equiv1/T$, 
the inverse temperature (We set $k_\mathrm{B}=\hbar=1$). The single particle Green's function $G$ and anomalous Green's function $F$ 
satisfy the Gor'kov equations:
\begin{subequations}
\begin{align*}
&&G(\mathbf{r},\tau;\mathbf{r}',\tau')=G_0(\mathbf{r},\tau;\mathbf{r}',\tau')-\frac{1}{16}\int_0^\beta \sum_{\mu,\nu}d\tau_1 d\tau_2\int d\mathbf{r}_1d\mathbf{r}_2 \left[G_0(\mathbf{r},\tau;\mathbf{r}_1,\tau_1)\Psi_\mu(\mathbf{r}_1+\frac{a\hat{\mu}}{2},\tau_1)\right. \nonumber \\
&&\left. G_0(\mathbf{r}_2+a\hat{\nu},\tau_2;\mathbf{r}_1+a\hat{\mu},\tau_1)\Psi^*_\nu(\mathbf{r}_2+\frac{a\hat{\nu}}{2},\tau_2)G(\mathbf{r}_2,\tau_2;\mathbf{r}',\tau')\right]~~~~~~~~~~~~~~~~~~~~~~\tag{S1a}\label{eq.GorkovEqn_G}\\
&&F(\mathbf{r},\mathbf{r}';\tau,\tau')=\frac{1}{4}\sum_\mu \int_0^\beta d\tau_1 \int d\mathbf{r}_1 G_0(\mathbf{r},\tau;\mathbf{r}_1,\tau_1)\Psi_\mu(\mathbf{r}_1+\frac{a\hat{\mu}}{2},\tau_1)G(\mathbf{r}',\tau';\mathbf{r}_1+a\hat{\mu},\tau')~~~~~~~~\tag{S1b}\label{eq.GorkovEqn_F}
\end{align*}
\end{subequations}
The indices $\mu,~\nu$ run over bond-directions $\pm\hat{x},\pm\hat{y}$. The factors $1/16$ and $1/4$ are related
to the normalization of $\Psi$. 

The self-energy shown in the inset of Fig.1(a) of the paper is obtained as follows.
Generalizing the s-wave approach of ref.~\cite{Stephen1992_S}, we 
first average the above equations over the 
configurations $\{\Psi_\mu(\mathbf{r},\tau)\}$ 
and then make the decoupling approximation
\begin{subequations}
\begin{align*}
&&\langle \Psi_\mu(\mathbf{r}_1+\frac{a\hat{\mu}}{2},\tau_1)\Psi^*_\nu(\mathbf{r}_2+\frac{a\hat{\nu}}{2},\tau_2)G(\mathbf{r}_2,\tau_2;\mathbf{r'},\tau')\rangle\approx\langle \Psi_\mu(\mathbf{r}_1+\frac{a\hat{\mu}}{2},\tau_1)\Psi^*_\nu(\mathbf{r}_2+\frac{a\hat{\nu}}{2},\tau_2)\rangle \langle G(\mathbf{r}_2,\tau_2;\mathbf{r'},\tau')\rangle\tag{S2a}\\
&&\langle \Psi_\mu(\mathbf{r}_1+\frac{a\hat{\mu}}{2},\tau_1)G(\mathbf{r}',\tau';\mathbf{r}_1+a\hat{\mu},\tau_1)\rangle \approx \langle \Psi_\mu(\mathbf{r}_1+\frac{a\hat{\mu}}{2},\tau_1)\rangle \langle G(\mathbf{r}',\tau';\mathbf{r}_1+a\hat{\mu},\tau_1)\rangle~~~~~~~~~~~~~\tag{S2b}
\end{align*} 
\end{subequations}
By definition $\langle \Psi_\mu(\mathbf{r},\tau)\rangle =0$ in a vortex liquid, which implies that 
$\langle F(\mathbf{r},\tau;\mathbf{r}',\tau')\rangle=0$.

For notational simplicity, we denote $\langle G(\mathbf{r},\tau;\mathbf{r}',\tau')\rangle$ by $G(\mathbf{r},\tau;\mathbf{r}',\tau')$. 
Eq.~(\ref{eq.GorkovEqn_G}) can then be written in the form
\begin{align*}
&&G(\mathbf{r},\mathbf{r}';i\omega_l)=G_0(\mathbf{r},\mathbf{r}';i\omega_l)-\frac{T}{16}\sum_m\sum_{\mu,\nu}e^{-i\Theta_{\mu\nu}}\int_{\mathbf{r}_1,\mathbf{r}_2}
\left[\exp[{-i\frac{2e}{c}\int_{\mathbf{r}_1+\frac{a\hat{\mu}}{2}}^{\mathbf{r}_2+\frac{a\hat{\nu}}{2}}\mathbf{A}\cdot d\mathbf{l}}]G_0(\mathbf{r},\mathbf{r}_1;i\omega_l)\right.\nonumber\\
&&\left. D_{\mu\nu}\left(\mathbf{r}_2-\mathbf{r}_1+\frac{a\hat{\nu}-a\hat{\mu}}{2},i\Omega_m\right)
 G_0(\mathbf{r}_2+a\hat{\nu},\mathbf{r}_1+a\hat{\mu};-i\omega_l+i\Omega_m)G(\mathbf{r}_2,\mathbf{r}';i\omega_l)\right]~~~~~\tag{S3} \label{eq.AveGorkovEqn}
\end{align*}
Here $\omega_l=(2l+1)\pi T$ and $\Omega_l=2l\pi T$ are Fermi and Bose Matsubara frequencies, respectively
and $\int_{\mathbf{r}_1}$ denotes $\int d\mathbf{r}_1$. Using the the separable form of the {\em gauge invariant} fluctuation propagator (Eq.~(1) of the paper), we find that  
\begin{align*}
D_{\mu\nu}(\mathbf{r},i\Omega_m)= s_{\mu\nu}\mathcal{D}(\mathbf{r})\mathcal{F}(i\Omega_m)
= s_{\mu\nu} \Delta^2\exp{(-r^2/2\ell^2)} \frac{1}{|\O_m|+\G}\tag{S4} \label{eq.PairPropagator_S}
\end{align*}
Note that $\mathcal{F}(i\Omega_m)$ is the Matsubara representation
of the dissipative form
$\mathrm{Im}\mathcal{F}^{(R)}(\Omega)=\Omega/(\Omega^2+\Gamma^2)$.

The electromagnetic phase factor appearing in Eq.~\eqref{eq.AveGorkovEqn} results from the definition of 
gauge-invariant propagator defined in main text. The integral $\int_\mathbf{0}^\mathbf{r}\mathbf{A}.d\mathbf{l}$ is taken along a straight line 
connecting $0$ and $\mathbf{r}$. The reasoning for this parallels that given in ref.~\cite{Stephen1992_S} for the s-wave case.
One additional complication here is the phase factor  
$\Theta_{\mu\nu}=(a/2\ell)^2(\nu_x+\mu_x)(\nu_y-\mu_y)$ arising from the 
definition of the d-wave pair field on the bonds of the lattice. 
(That this is a lattice related effect is evident from the fact that $\Theta_{\mu\nu}$ is negligible in the low field limit, $a\ll\ell$.) 

We choose to work in the Landau gauge $\mathbf{A}=Hx\hat{y}$ to represent the magnetic field $H\hat{z}$.
For a parabolic dispersion the Green's functions in the Landau level (LL) basis is
$G(nq,n'q';i\omega_l)=\int d\mathbf{r}d\mathbf{r}' \phi^*_{nq}(\mathbf{r})\phi_{n'q'}(\mathbf{r}')G(\mathbf{r},\mathbf{r}';i\omega_l)$.  Here $\phi_{nq}(\mathbf{r})$ is the LL wave function with $n$ the LL index and $q$ going over the degenerate
states in each LL. We can now rewrite Eq.\eqref{eq.AveGorkovEqn} in the LL basis so that
it has the form of Dyson's equation $G=G_0+G_0\Sigma G$.
The bare $G_0(nq,n'q';i\omega_l)\equiv G_0(n,i\omega_l)\delta_{nn'}\delta_{qq'}$ 
with $G_0(n,i\omega_l)=(i\omega_l-\xi_n)^{-1}$. The self-energy is given by
\begin{align*}
&&\Sigma(nq,n'q';i\omega_l)=\frac{T}{4}\left(\frac{\Delta}{2}\right)^2\sum_{m}\sum_{\mu,\nu}\sum_{n_1}s_{\mu\nu}
I_{\mu\nu}(nq,n'q',n_1)\frac{\mathcal{F}(i\Omega_m)}{i\omega_l-i\Omega_m+\xi_{n_1}};\tag{S5} \label{eq.Self-energy_general}
\end{align*}
\begin{eqnarray*}
&&I_{\mu\nu}(nq,n'q',n_1)\\
&&=e^{-i\Theta_{\mu\nu}}\sum_{q_1}\int_{\mathbf{r}_1,\mathbf{r}_2}\exp[{-i\frac{2e}{c}\int_{\mathbf{r}_1+\frac{a\hat{\mu}}{2}}^{\mathbf{r}_2+\frac{a\hat{\nu}}{2}}\mathbf{A}\cdot d\mathbf{l}}]\mathcal{D}\left(\mathbf{r}_2-\mathbf{r}_1+\frac{a\hat{\nu}-a\hat{\mu}}{2}\right)\phi_{nq}^*(\mathbf{r}_1)\phi_{n_1q_1}(\mathbf{r}_2+a\hat{\nu})\phi_{n_1q_1}^*(\mathbf{r}_1+a\hat{\mu})\phi_{n'q'}(\mathbf{r}_2) \nonumber
\end{eqnarray*}
The Matsubara sum can be done in the dynamic case ($\Gamma\neq 0$),
using the form of the $\mathcal{F}(i\Omega_m)$ in Eq.\eqref{eq.PairPropagator_S}. 
In the static approximation $D_{\mu\nu}(\mathbf{r},\tau)$ is independent of $\tau$ so that
$\mathcal{F}(i\Omega_m)=\delta_{m,0}/T$. As a result, the Matsubara sum in Eq.\eqref{eq.Self-energy_general} 
is trivial. We discuss below how the static limit is recovered by 
taking a suitable high-temperature or {\it classical} limit of the dynamic case.

We can do the $\sum_{q_1}\int_{\mathbf{r}_1,\mathbf{r}_2}$ integrals and
express $I_{\mu\nu}(nq,n'q',n_1)$ in terms of special functions, with
the various phase factors ``canceling out''.
We omit the details of this lengthy algebra here. 
The self-energy evaluated from Eq.~\eqref{eq.Self-energy_general} turns out to be diagonal in $q$-space.
It is also independent of $q$, which can be traced to the fact that the vortex-liquid state does not break translational invariance.
We denote it by $\Sigma(n,n';i\omega_l)$ in the main text. Unlike the $s$-wave case~\cite{Stephen1992_S}, 
$\Sigma$ is a matrix in the LL index due to the $d$-wave nature of the order parameter. We can show that 
the matrix elements $\Sigma(n,n';i\omega_l)$ decay rapidly away from the diagonal $n=n'$ like 
$\sim\exp{[-(n-n')^2/4n]}$. In the spirit of our non-self-consistent calculation [Fig.1a (inset)], 
given that $G_0$ is diagonal in the LL index, we only retain the dominant diagonal terms in 
$\Sigma(n,n;i\omega_l)\equiv \Sigma(n,i\omega_l)$. The ${\bf k}$-space analysis
described below serves to validate this approximation.

In the dynamic case, the imaginary part of the retarded self-energy ($i\omega_l\rightarrow \omega+i0^+$) is
\begin{align*}
&&\mathrm{Im}\Sigma(n,\omega)=\left(\frac{\Delta}{2}\right)^2\sum_{n_1}\left\{ I_{nn_1}\left[1+e^{-\frac{a^2}{2\ell^2}}
L_{n+n_1}\left(\frac{a^2}{\ell^2}\right)-2e^{-\frac{a^2}{4\ell^2}}L_{n+n_1}\left(\frac{a^2}{2\ell^2}\right)\right]\right.\nonumber \\
&&\left.\int\frac{d\Omega}{2\pi}\left[\coth\left(\frac{\Omega}{2T}\right)-\tanh\left(\frac{\Omega-\omega}{2T}\right)\right]\mathrm{Im}G_0(n_1,\Omega-\omega)\mathrm{Im}\mathcal{F}^{(R)}(\Omega)\right\} \tag{S6}\label{eq.Self-energy_exact}
\end{align*}
Here $I_{nn_1}=(n+n_1)!/n!n_1!2^{n+n_1+1}$ and $L_n(x)$ is a Laguerre polynomial. The real part of the self-energy can be obtained using Kramers-Kronig relation from $\mathrm{Im}\Sigma(n,\omega)$. 
For large values of the LL index $n$ we can derive a useful approximation 
\begin{align*}
&&\mathrm{Im}\Sigma(n,\omega)\simeq-\left(\frac{\Delta}{2}\right)^2\left[1+
J_0\left(2\sqrt{2n}\frac{a}{\ell}\right)-2J_0\left(2\sqrt{n}\frac{a}{\ell}\right)\right] 
\left\{\frac{1}{4\sqrt{n\pi}\hbar\omega_c}\int d\Omega\left[\coth\left(\frac{\Omega}{2T}\right)\right.\right.\nonumber\\
&&\left.\left.-\tanh\left(\frac{\Omega-\omega}{2T}\right)\right]\frac{\Omega}{\Omega^2+\Gamma^2}e^{-\frac{(\omega-\Omega+\xi_n)^2}{4n\hbar^2\omega_c^2}}\right\}~~~~~~~~~~~~~~~~~~~~~~~~~~~~~~~~~~~~~\tag{S7}\label{eq.Self-energy_approx}
\end{align*}
where $J_0(x)$ denotes a Bessel function. We have benchmarked this form 
by comparing it with results obtained directly from Eq.\eqref{eq.Self-energy_exact} 
and find that the approximation is accurate for $n_\mathrm{F}=(\mu/\hbar \omega_c)\gtrsim 10$. We 
have used Eq.\eqref{eq.Self-energy_approx} for computing the results reported in the main paper. 
We have also found expressions analogous to Eq.\eqref{eq.Self-energy_exact} and Eq.\eqref{eq.Self-energy_approx} 
for the case of static phase fluctuations separately, and summarized in the Methods section.

\subsection{$\mathbf{k}$-space calculation}

We now discuss the $\mathbf{k}$-space calculation of self-energy $\Sigma(\mathbf{k},\omega)$ used in the main text
together with a semiclassical approach to quantum oscillations for $\mu>\Delta\gg\hbar \omega_c$. 
As a result of the translational invariance of the vortex-liquid state, one can make a gauge transformation and 
represent the (non-gauge-invariant) Green's function $G(\mathbf{r},\mathbf{r}';i\omega_l)$ as
\begin{align*}
&&G(\mathbf{r},\mathbf{r}';i\omega_l)=\widetilde{G}(\mathbf{r}'-\mathbf{r},i\omega_l)\exp[{-i\frac{e}{c}\int_\mathbf{r}^{\mathbf{r}'}\mathbf{A}.d\mathbf{l}}],~~~~~~~~~~~~~~~\tag{S8}
\end{align*}    
where $\widetilde{G}$ depends only on the separation $\mathbf{r}'-\mathbf{r}$. 
We rewrite Eq.\eqref{eq.AveGorkovEqn} in terms of $\widetilde{G}(\mathbf{r},i\omega_l)$ or its Fourier transform $G(\mathbf{k},i\omega_l)$
(where we omit the tilde to simplify notation) and obtain the corresponding self-energy 
\begin{align*}
&&\Sigma(\mathbf{k},i\omega_l)=-T\int\frac{d^2q}{(2\pi)^2}\sum_m \mathcal{D}(\mathbf{q}) \mathcal{F}(i\Omega_m)
G_0(\mathbf{k}-\mathbf{q},-i\omega_l+i\Omega_m)\left[\cos{\left(k_xa-\frac{q_xa}{2}\right)}-\cos{\left(k_ya-\frac{q_ya}{2}\right)}\right]^2. \tag{S9}
\end{align*}
Here $\mathcal{D}(\mathbf{q}) = \exp(-q^2\ell^2/2)$ is the Fourier transform of $\mathcal{D}(\mathbf{r})$  in Eq.\eqref{eq.PairPropagator}. 

In the semiclassical regime where $\mu \gg \hbar\omega_c$ or the cyclotron radius $R_c \gg \ell$ the inter-vortex separation, we neglect the effects of quantization of electronic orbits while evaluating $\Sigma(\mathbf{k},i\omega_l)$ and approximate 
$G_0(\mathbf{k},i\omega_l)$ by $(i\omega_l-\xi_\mathbf{k})^{-1}$. We can further simplify this result by noting 
that in the field regime of interest $\ell \gg \xi_0$ and $\mathcal{D}(\mathbf{q})$ is sharply peaked around $\mathbf{q}=0$ 
with a width $\sim 1/\ell$. This enables us to make the expansions
$\cos{(k_xa-\frac{q_xa}{2})}\simeq \cos(k_xa)$ and $\xi_{\mathbf{k}-\mathbf{q}}\simeq \xi_\mathbf{k}-\mathbf{v}_\mathbf{k}.\mathbf{q}$, where
$\mathbf{v}_\mathbf{k}=(\partial\xi_\mathbf{k}/\partial \mathbf{k})$. Proceeding in a manner similar to
the LL discussion above, we obtain for dynamical fluctuation case
\begin{align*}
&&\mathrm{Im}\Sigma(\mathbf{k},\omega)=-\frac{\Delta_\mathbf{k}^2\ell}{2\sqrt{2\pi}v_\mathbf{k}}\int d\Omega\left[\coth{\left(\frac{\Omega}{2T}\right)}-\tanh{\left(\frac{\Omega-\omega}{2T}\right)}\right]\frac{\Omega}{\Omega^2+\Gamma^2}e^{-\frac{(\omega-\Omega+\xi_\mathbf{k})^2\ell^2}{2v_\mathrm{k}^2}}\tag{S10}\label{eq.ImSigma_kSpace}
\end{align*} 
Here $\Delta_\mathbf{k}=(\Delta/2)(\cos{k_xa}-\cos{k_ya})$ and $v_\mathbf{k}=|\mathbf{v}_\mathbf{k}|$. 
The real part of $\Sigma(\mathbf{k},\omega)$ can be obtained by Kramers-Kronig relation.
The corresponding expression for $\Sigma(\mathbf{k},\omega)$ in the static case is shown in the Methods.
 
\subsection{The high-temperature (classical) limit and the static case}

As noted in the paper, one cannot just take the (singular) $\Gamma \rightarrow 0$ limit of the dynamic phase fluctuations
at low temperatures to describe static phase fluctuations.
Here we show the relation between the high-temperature limit of the self-energy of Eq.\eqref{eq.ImSigma_kSpace} for the dynamic case 
and static case self-energy of Eq.~(2) of the Methods section. 
In the high temperature limit $T\gg\Gamma$, we see that
 $T\gg\Omega$ for the frequencies contributing to the integral of Eq.\eqref{eq.ImSigma_kSpace}. 
Using $\coth(\Omega/2T)\simeq 2T/\Omega$ and $\tanh(\Omega/2T) \simeq \Omega/2T$ we get
\begin{align*}
&&\mathrm{Im}\Sigma(\mathbf{k},\omega)\simeq \frac{T\Delta_\mathbf{k}^2\ell}{\sqrt{2\pi}v_\mathbf{k}}\int d\Omega \frac{1}{\Omega^2+\Gamma^2}e^{-\frac{(\omega+\xi_\mathbf{k}-\Omega)^2\ell^2}{2v_\mathbf{k}^2}}~~~~~~~~~~~\tag{S11}
\end{align*}  
To see that this is qualitatively similar to the self-energy for the static case  
we approximate the Lorentzian $1/(\Omega^2+\Gamma^2)$ above
by a Gaussian $\sqrt{\pi/2}\exp(-\Omega^2/2\Gamma^2)/\Gamma^2$. We thus obtain 
\begin{align*}
&&\mathrm{Im}\Sigma(\mathbf{k},\omega)\approx \left(\frac{T\Delta_\mathbf{k}^2}{\Gamma}\right) \sqrt{\frac{\pi}{2}} \frac{\ell}{\sqrt{v_\mathbf{k}^2+\Gamma^2\ell^2}}e^{-\frac{(\omega+\xi_\mathbf{k})^2\ell^2}{2(v_\mathbf{k}^2+\Gamma^2\ell^2)}}~~~~~~~~~\tag{S12}
\end{align*}
This has exactly the same form as the static case (Eq.~(2) of Methods)
with some redefinitions:  $(T\Delta^2/\Gamma)$ going to $\Delta^2$ and $\sqrt{v_\mathbf{k}^2+\Gamma^2\ell^2}$ to $v_\mathbf{k}$. 
For a closely related discussion, see ref.~\cite{Micklitz2009_S}.

\subsection{DOS in the low field regime: $\sqrt{H}$ plus impurity scattering}

To investigate the asymptotic behavior of the DOS, $N_H(0)$ as a function of field and impurity scattering strength,
we focus on the self-energy for the static case. To obtain analytical results, we approximate the Gaussian in 
Eq.~(4) of Methods by a Lorentzian, so that we get

\begin{align*}
&&\Sigma(\mathbf{k},\omega)\approx \frac{\Delta_\mathbf{k}^2}{\omega+\xi_\mathbf{k}+i(\gamma_0+v_\mathbf{k}/\ell)}-i\gamma_0~~~~~~~~~~~~~~~~~\tag{S13}
\label{impurity_se}
\end{align*} 
 
where $\gamma_0 \ll \Delta \leq \mu$ comes from weak impurity scattering. 
This corresponds to a static pair correlator $\mathcal{D}(r)\sim \exp(-r/\ell)$ (that decays exponentially rather than as a Gaussian)  and gives rise to qualitatively similar low-energy electronic spectra and DOS. We can also use a more general form of \eqref{impurity_se}
with $\gamma(\omega)$, but only its $\omega=0$ limit $\gamma_0$ is relevant below.

We calculate $N_H(0)$ as
\begin{align*}
&&N(0)=-\frac{2}{\pi}\int \frac{d^2k}{(2\pi)^2}\mathrm{Im}G(\mathbf{k},0)=\frac{1}{2\pi^3}\int d^2k \frac{(\gamma_0+v_\mathbf{k}/\ell)(\xi_\mathbf{k}^2+\Delta_\mathbf{k}^2+\gamma_0^2+\gamma_0v_\mathbf{k}/\ell)-\xi_\mathbf{k}^2v_\mathbf{k}/\ell}{(\xi_\mathbf{k}^2+\Delta_\mathbf{k}^2+\gamma_0^2+\gamma_0v_\mathbf{k}/\ell)^2+\xi_\mathbf{k}^2(v_\mathbf{k}/\ell)^2}
\tag{S14}
\end{align*} 
We next expand $\Delta_\mathbf{k}\approx v_\Delta p_\parallel$ and $\xi_\mathbf{k}\approx v_\mathrm{F}p_\perp$, with $\mathbf{p}=\mathbf{k}-\mathbf{k}_N$, in the same way near the nodes $\mathbf{k}_N$'s as described in Methods (where we did not include
impurity scattering). In the low-field regime $H\rightarrow 0$, we obtain $N_H(0)$ by expanding in powers of $(1/\ell)$ to get
\begin{align*}
&&N_H(0)\approx\frac{2}{\pi^3}\left[\gamma_0\int_{-p_0}^{p_0} dp_\perp dp_\parallel \frac{1}{v_\mathrm{F}^2p_\perp^2+v_\Delta^2p_\parallel^2+\gamma_0^2}+\frac{v_\mathrm{F}}{\ell}\int_{-p_0}^{p_0} dp_\perp dp_\parallel \frac{v_\Delta^2p_\parallel^2}{(v_\mathrm{F}^2p_\perp^2+v_\Delta^2p_\parallel^2+\gamma_0^2)^2}\right],
\tag{S15}\label{eq.N_H(0)}
\end{align*}
where the momentum cut-off $p_0\simeq \pi/a$. This implies that $N_H(0)\approx A \gamma_0+ B \sqrt{H}$ as $H\rightarrow 0$, where $A$ and $B$ are appropriate constants that depend logarithmically on $\gamma_0$.

\end{widetext}


\begin{thebibliography}{99} 
\bibitem{Leyraud2007} Doiron-Leyraud, N.~et al. Quantum oscillations and the Fermi surface in an underdoped high-$T_c$ superconductor. {\it Nature} {\bf 447}, 565-568 (2007).
\bibitem{Sebastian2008} Sebastian, S.~E.~et al. A multi-component Fermi surface in the vortex state of an underdoped high-$T_c$ superconductor. {\it Nature} {\bf 454}, 200-203 (2008). 
\bibitem{Riggs2011} Riggs, S.~C.~et al. Heat capacity through the magnetic-field-induced
resistive transition in an underdoped high-temperature superconductor. {\it Nature Phys.}~{\bf 7}, 332-335 (2011).
\bibitem{Timusk1999} Timusk, T.~\& Statt, B. The pseudogap in high-temperature superconductors: an experimental survey. {\it Rep.~Prog.~Phys.}~{\bf 62}, 61-122 (1999). 
\bibitem{Kanigel2006} Kanigel, A.~et al. Evolution of the pseudogap from Fermi arcs to the nodal liquid. {\it Nature Phys.}~2, 447-451 (2006). 
\bibitem{Volovik1993} Volovik, G.~E. Superconductivity with lines of gap nodes: density of states in the vortex. {\it JETP Lett.} {\bf 58}, 469-473 (1993).
\bibitem{Loram1993} Loram, J.~W., Mirza, K.~A., Cooper, J.~R.~\& Liang, W.~Y. Electronic specific heat of $\mathrm{YBa_2Cu_3O_{6+x}}$ from 1.8 to 300 K. {\it Phys.~Rev.~Lett.}~{\bf 71}, 1740–1743 (1993). 
\bibitem{Li2010} Li, L.~et al. Diamagnetism and Cooper pairing above $T_c$ in cuprates. {\it Phys.~Rev.~B} {\bf 81}, 054510-(1-9) (2010).
\bibitem{Wang2006} Wang, Y., Li, L.~\& Ong, N.~P. Nernst effect in high-$T_c$ superconductors. {\it Phys.~Rev.~B} {\bf 73}, 024510-(1-20) (2006). 
\bibitem{Uemura1989} Uemura, Y.~J.~et al. Universal Correlations between $T_c$ and $n_s/m^*$ (Carrier Density over Effective Mass) in High-$T_c$ Cuprate Superconductors. {\it Phys.~Rev.~Lett.}~{\bf 62}, 2317–2320 (1989).
\bibitem{Emery1994} Emery, V.~J.~\& Kivelson, S.~A. Importance of phase fluctuations in superconductors with small superfluid density. {\it Nature} {\bf 374}, 434-437 (1994).
\bibitem{Hetel2008} Hetel, I., Lemberger, T.~R.~\& Randeria, M. Quantum critical behaviour in the superfluid density of strongly underdoped ultrathin copper oxide films. {\it Nature Phys.}~{\bf 3}, 700-702 (2007). 
\bibitem{Maki1991} Maki, K. Quantum oscillation in vortex states of type-II superconductors. {\it Phys.~Rev.~B} {\bf 44}, 2861-2862 (1991).
\bibitem{Stephen1992} Stephen, M.~J. Superconductors in strong magnetic fields: de Haas-van Alphen effect. {\it Phys.~Rev.~B} {\bf 45}, 5481–5485 (1992). 
\bibitem{Maniv2001} Maniv, T., Zhuravlev, V., Vagner, I.~\& Wyder, P. Vortex states and quantum magnetic oscillations in conventional type-II superconductors. {\it Rev.~Mod.~Phys.}~{\bf 73}, 867–911 (2001). 
\bibitem{Vignolle2008} Vignolle, B.~et al. Quantum oscillations in an overdoped high-$T_c$
superconductor. {\it Nature} {\bf 455}, 952-955 (2008).
\bibitem{Millis2007} Millis, A.~J.~\& Norman, M.~R. Antiphase stripe order as the origin of electron pockets observed in 1/8-hole-doped cuprates. {\it Phys.~Rev.~B} {\bf 76}, 220503(R)-(1-4) (2007).
\bibitem{Chakravarty2008} Chakravarty, S.~\& Kee, H.-Y. Fermi pockets and quantum oscillations of the Hall coefficient in high-temperature superconductors. {\it Proc.~Natl.~Acad.~Sci.~USA} {\bf 105}, 8835-8839 (2008).
\bibitem{Yao2011} Yao, H., Lee, D.-H.~\& Kivelson, S. Fermi-surface reconstruction in a smectic phase of a high-temperature superconductor. {\it Phys.~Rev.~B} {\bf 84}, 012507-(1-5) (2011).
\bibitem{Harrison2011} Harrison, N.~\& Sebastian, S.~E. Protected Nodal Electron Pocket from Multiple-Q Ordering in Underdoped High Temperature Superconductors. {\it Phys.~Rev.~Lett.}~{\bf 106}, 226402-(1-4) (2011).
\bibitem{Sebastian2011} Sebastian, S.~E., Harrison, N.~\& Lonzarich, G.~G. Quantum oscillations in the high-$T_c$ cuprates. {\it Phil.~Trans.~R.~Soc.~A} {\bf 369}, 1687-1711 (2011).
\bibitem{Senthil2009} Senthil, T.~\& Lee, P.~A. Synthesis of the phenomenology of the underdoped cuprates. {\it Phys.~Rev.~B} {\bf 79}, 245116-(1-7) (2009).
\bibitem{Micklitz2009} Micklitz, T.~\& Norman, M.~R. Nature of spectral gaps due to pair formation in superconductors. {\it Phys.~Rev.~B} {\bf 80}, 220513(R)-(1-4) (2009).
\bibitem{Banerjee2011} Banerjee, S., Ramakrishnan, T.~V.~\& Dasgupta, C. Effect of pairing fluctuations on low-energy electronic spectra in cuprate superconductors. {\it Phys.~Rev.~B} {\bf 84}, 144525-(1-12) (2011).
\bibitem{Luttinger1961} Luttinger, J.~M. Theory of the de Haas-van Alphen Effect for a System of Interacting Fermions. {\it Phys.~Rev.}~{\bf 121}, 1251–1258 (1961).  
\bibitem{Wu2011} Wu, T.~et al. Magnetic-field-induced charge-stripe order in the high-temperature superconductor $\mathrm{YBa_2Cu_3O_y}$. {\it Nature} {\bf 477}, 191-194 (2011).    
\end{thebibliography}

\begin{thebibliography}{99} 

\bibitem{Stephen1992_S} Stephen, M.~J. Superconductors in strong magnetic fields: de Haas-van Alphen effect. {\it Phys.~Rev.~B} {\bf 45}, 5481 (1992). 
\bibitem{Micklitz2009_S} Micklitz, T.~\& Norman, M.~R. Nature of spectral gaps due to pair formation in superconductors. {\it Phys.~Rev.~B} {\bf 80}, 220513(R)-(1-4) (2009).
\end{thebibliography}
\end{document}